\documentclass[10pt,conference]{IEEEtran}

\usepackage{times}
\usepackage{graphicx}
\usepackage{subfigure}
\usepackage{array} % used for tabular environments
\newcolumntype{P}[1]{>{\centering\arraybackslash}p{#1}}
\usepackage{amsmath,amsfonts,amssymb,amsthm}
\DeclareMathOperator*{\argminA}{arg\,min}
\DeclareMathOperator*{\argmaxA}{arg\,max}
\usepackage{color}
\usepackage{url}
\usepackage{listings} % used for source code
\usepackage{xspace} % package that may or may not insert "eaten" spaces.
\usepackage{booktabs} % optimizations for tables
\usepackage{hyperref} % hyperlinks
\usepackage[noabbrev]{cleveref}
\usepackage[font=small]{caption}
\usepackage{courier} % proper bold version
\usepackage{verbatim}
\usepackage{xcolor}

\usepackage{enumitem}

\usepackage{calc}
\usepackage[linesnumbered,lined,ruled,vlined]{algorithm2e}
\usepackage{multirow}

\let\oldenumerate\enumerate
\renewcommand{\enumerate}{
  \oldenumerate
  \setlength{\itemsep}{.0pt}
  \setlength{\parskip}{0pt}
  \setlength{\parsep}{0pt}
}

\let\olditemize\itemize
\renewcommand{\itemize}{
  \olditemize
  \setlength{\itemsep}{1pt}
  \setlength{\parskip}{0pt}
  \setlength{\parsep}{0pt}
}

\newtheorem*{proposition*}{Proposition}
\theoremstyle{definition}

\newcommand{\ie}{{\em i.e.,~}}
\newcommand{\eg}{{\em e.g.,~}}

\def\F{Fig.~}

\def\S{Section~}
\renewcommand{\figurename}{Fig.}

\newcommand{\ar}[3]{} %% something bibtex is doing - ignore it
\include{latexdefs}
\include{mathnotation}

\newcommand{\heading}[1]{{\vspace{0pt}\noindent\bf{#1}}} % inside section
{\makeatletter
 \gdef\xxxmark{%
   \expandafter\ifx\csname @mpargs\endcsname\relax % in minipage?
     \expandafter\ifx\csname @captype\endcsname\relax % in figure/caption?
       \marginpar{\textcolor{red}{xxx~}}% not in a caption or minipage, can use marginpar
     \else
       \textcolor{red}{xxx~}% notice trailing space
     \fi
   \else
     \textcolor{red}{xxx~}% notice trailing space
   \fi}
 \gdef\xxx{\@ifnextchar[\xxx@lab\xxx@nolab}
 \long\gdef\xxx@lab[#1]#2{{\bf [\xxxmark \textcolor{red}{#2} ---{\sc #1}]}}
 \long\gdef\xxx@nolab#1{{\bf [\xxxmark \textcolor{red}{#1}]}}
}

{\makeatletter
 \gdef\edit{\@ifnextchar[\edit@lab\edit@nolab}
 \long\gdef\edit@lab[#1]#2{[\textcolor{red}{#2} ---{\sc #1}]}
 \long\gdef\edit@nolab#1{[\textcolor{red}{#1}]}
}

\newcommand{\ignore}[1]{}

\definecolor{grey}{rgb}{0.5,0.5,0.5}

\lstdefinelanguage{yaml}{
  alsoletter=/,
  ndkeywords={id, post, body, string, /blog/posts/, /api, string,
              integer},
  ndkeywordstyle=\color{purple},
  basicstyle=\footnotesize,
}

\lstdefinelanguage{restler}{
  keywords={requests, Request, set_var, writer, dependencies, parser,
            post_send, restler_static, restler_fuzzable},
  keywordstyle=\color{darkgray}\bfseries,
  stringstyle=\color{purple},
  morestring=[b]",
  basicstyle=\footnotesize,
}

\lstdefinelanguage{networklog}{
  alsoletter={\\, @, .},
  ndkeywords={
      /api/projects,
      /api/projects/1243/repository/branches,
      /api/projects/1243/repository/commits,
      /api/projects/1243/repository/commits/8a53ed/cherry_pick,
      POST, 201, Created, 500, Internal, Server, Error,  HTTP/1.1},
  ndkeywordstyle=\color{black}\bfseries,
  morekeywords={internal, server, error, admin\\xd7@example.com},
  keywordstyle=\color{red}\bfseries,
  basicstyle=\footnotesize,
}

\lstdefinelanguage{bucketlog}{
  alsoletter=/0123456789.,
  ndkeywords={HTTP/1.1, string, /api/v4/projects/projectid/repository/commits,
              /api/v4/projects, POST},
  ndkeywordstyle=\color{purple},
  keywords={requests, Request, set_var, writer, dependencies, parser,
            post_send, restler_static, restler_fuzzable},
  keywordstyle=\color{darkgray}\bfseries,
  basicstyle=\footnotesize,
}

\def\pythia{Pythia\xspace}
\def\restler{RESTler\xspace}
\def\gitlab{GitLab\xspace}
\def\mastodon{Mastodon\xspace}
\def\nbugs{29\xspace}

\begin{document}
\title{Pythia: Grammar-Based Fuzzing of REST APIs with Coverage-guided Feedback and Learning-based Mutations}
\author{Vaggelis Atlidakis\\Columbia University \and Roxana Geambasu\\Columbia University \and Patrice Godefroid\\Microsoft Research \and Marina Polishchuk\\Microsoft Research \and Baishakhi Ray\\Columbia University}
\maketitle

\maketitle
\thispagestyle{plain}
\pagestyle{plain}
\begin{abstract}
This paper introduces \pythia, the first fuzzer that augments grammar-based
fuzzing with coverage-guided feedback and a learning-based mutation strategy for
stateful REST API fuzzing. \pythia uses a statistical model to learn
common usage patterns of a target REST API from structurally valid
seed inputs. It then generates learning-based mutations by
injecting a small amount of noise deviating from common usage patterns
while still maintaining syntactic validity. \pythia's
mutation strategy helps generate grammatically valid test cases and
coverage-guided feedback helps prioritize the test cases that are
more likely to find bugs. We present experimental evaluation
on three production-scale, open-source cloud services showing
that \pythia outperforms prior approaches both in code coverage and
new bugs found. Using \pythia, we found \nbugs new
bugs which we are in the process of reporting to the respective service owners.
\end{abstract}

\section{Introduction}
\label{sec:introduction}

Fuzzing~\cite{fuzzing-book} is a popular approach to find bugs in
software.  It involves generating new test inputs and feeding them to
a target application which is continuously monitored for errors. Due
to its simplicity, fuzzing has been widely adopted and has found
numerous security and reliability bugs in many real-world
applications. At a high level, there are three main approaches to
fuzzing~\cite{God20}: blackbox random fuzzing, grammar-based fuzzing,
and whitebox fuzzing.

Blackbox random fuzzing simply randomly mutates well-formed program
inputs and then runs the program with those mutated inputs with the
hope of triggering bugs. This process can be guided by code-coverage
feedback which favors the mutations of test inputs that exercize new
program statements~\cite{AFL}. Whitebox fuzzing~\cite{SAGE} can
further improve test-generation precision by leveraging more
sophisticated techniques like dynamic symbolic execution, constraint
generation and solving, but at a higher engineering cost. All these
blackbox, greybox, or whitebox {\em mutation-based fuzzing} techniques
work well when fuzzing applications with relatively simple input
binary formats, such as audio, image, or video processing
applications~\cite{jpeg,mp3,mp4}, ELF parsers~\cite{elf}, and other
binary utilities~\cite{binutils}.

However, when fuzzing applications with complex structured non-binary
input formats, such as XML parsers~\cite{xml}, language compilers or
interpreters~\cite{clang,gcc,python}, and cloud service
APIs~\cite{azure-apis}, their effectiveness is typically limited, and
{\em grammar-based fuzzing} is then a better alternative. With this
approach, the user provides an input grammar specifying the input
format, and may also specify what input parts are to be fuzzed and
how~\cite{Peach,SPIKE,boofuzz,burp}.  From such an input grammar, a
grammar-based fuzzer then generates many new inputs, each satisfying
the constraints encoded by the grammar. Such new inputs can reach
deeper application states and find bugs beyond syntactic lexers and
semantic checkers.

Grammar-based fuzzing has recently been {\em automated} in the domain
of REST APIs by \restler~\cite{restler}. Most production-scale cloud
services are programmatically accessed through REST APIs that are
documented using API specifications, such as OpenAPI~\cite{swagger}.
Given such a REST API specification, \restler automatically generates
a fuzzing grammar for REST API testing. \restler performs a
lightweight static analysis of the API specification in order to infer
dependencies among request types, and then automatically generates an
input grammar that encodes {\em sequences} of requests (instead of
single requests) in order to exercise the service behind the API more
deeply, in a {\em stateful manner}. However, the generated grammar
rules usually include few values for each primitive type, like strings
and numeric values, in order to limit an inevitable combinatorial
explosion in the number of possible fuzzing rules and values.  These
primitive-type values are either obtained from the API specification
itself or from a user-defined dictionary of values. All these values
remain static over time, and are not prioritized in any way. These
limitations (fuzzing rules with predefined sets of values and lack of
feedback) are typical in grammar-based fuzzing in general, beyond REST
API fuzzing.

To address these limitations, we introdude Pythia
\footnote{
    \pythia was an ancient Greek priestess who served as oracle,
    commonly known as the Oracle of Delphi, and was credited for various
    prophecies.
},
a new fuzzer that {\em augments grammar-based fuzzing with coverage-guided feedback
and a learning-based mutation strategy for stateful REST API fuzzing}. \pythia's
mutation strategy helps generate grammatically valid test cases and
coverage-guided feedback helps prioritize the test cases that are
more likely to find bugs.
This paper makes the following contributions:
\begin{itemize}
	\vspace{-3pt}
    \item We introduce \pythia, a new fuzzer that augments
        grammar-based fuzzing with coverage-guided feedback.
    \item We implement a learning-based mutation strategy for
        Stateful REST API Fuzzing.
    \item We present experimental evidence
        showing that by combining its learning-based mutation
        strategy and coverage-guided feedback, \pythia significantly
        outperforms prior approaches.
    \item We use \pythia to test three productions-scale, open-source
        cloud services (namely \gitlab, \mastodon, and Spree)
        with REST APIs specifying more than 200 request types.
    \item We discover new bugs in all three services tested so far. In total,
        we found \nbugs new bugs and we discuss several of these.
	\vspace{-3pt}
\end{itemize}
The rest of the paper is organized as follows. \Cref{sec:motivation} presents
background information on REST API fuzzing and the motivation
for this work. \Cref{sec:pythia} presents the design of
Pythia. \Cref{sec:subj,sec:results} presents experimental results
on three production-scale, open-source cloud services.
\Cref{sec:case-studies} discusses new bugs found
by \pythia. \Cref{sec:related-work,sec:conclusion} discuss related work
and conclusions.

\section{Background and Motivation}
\label{sec:motivation}

This paper aims at
testing cloud services
accessible through REpresentational State Transfer (REST) Application
Programming Interfaces (APIs)~\cite{rest}. A REST API is a finite set of
requests, where a request $r$ is a tuple $\langle t, p, h, b \rangle$,
as shown below. % (see~\Cref{tab:apidesc}).

\vspace{3pt}
{
% \begin{table}[h!tbp]
 \scriptsize
   \centering
%   \caption{REST API Description}
    \begin{tabular}{lp{0.7\linewidth}l}
      \toprule
    \textbf{Field} & \textbf{Description}\\
\midrule
    Request Type (t) & One of POST (create), PUT (create or update), GET (read),
        DELETE (delete), and PATCH (update).  \\
    Resource Path (p) & A string identifying a cloud resource and its parent
        hierarchy with the respective resource types and their names.    \\
    Header (h) &  Auxilary information about the requested entity. \\
    Body (b) &    Optional dictionary of data for the request to be executed successfully.  \\
\bottomrule
    \end{tabular}%
%   \label{tab:apidesc}%
%   \vspace{-15pt}
% \end{table}%
}
\vspace{3pt}

Consecutive REST API requests often have inter-dependencies w.r.t. some resources. For example, a request whose execution creates a new resource of type $T$
is called a {\em producer} of %for the resource type
$T$ and a request which requires $T$ in its path or body is called a {\em consumer} of $T$. A {\em producer-consumer} relationship between two requests is called a {\em dependency}.
The goal of our fuzzer, which is a client program, is to test a target service through the APIs. The fuzzer automatically generates and executes (i.e., sends) various API requests with the hope of triggering unexpected, erroneous behaviours.
We use the term {\em test case} to refer to
a sequence of API requests and the respective responses.
%, as shown in~\Cref{fig:testcase}.

\heading{Example REST API test case and detected bug.}
\Cref{fig:testcase} shows a sample \pythia test case for \gitlab\cite{gitlab},
an open-source cloud service for self-hosted repository management.
The test case contains three request-response pairs and exercises
functionality related to version control commit operations.
%The first line of each request and response are emphasized with bold for
%clarity.
The first request (\Cref{req:1}) POST %request of the sequence
creates a new \gitlab project.
It has a path without any resources and a body with a
dictionary of a non-optional parameter specifying the desired name of the
requested project (``21a8fa''). In response, it
receives back metadata describing the newly created project, including its unique id (\Cref{res:1}).
The second request, also of type POST, creates a repository branch in an existing project (\Cref{req:2}). It has a path specifying the previously created resource of type ``project'' and id ``1243'', and a body with a  parameter specifying the branch name (``feature1''), such that, the branch can be created within the previously created project.
In response (\Cref{res:2}), it receives back
a dictionary of metadata describing the newly created branch, including its designated name.
Finally, the last request (\Cref{req:3}) uses the latest branch (in its path) as well as the unique project id (in its body) and attempts to create a new commit.
The body of this request contains a set of parameters specifying the name of the existing target branch, the desired commit message
(``testString''), and the actions related to the new commit (i.e.,  creation of a file). However, the relative path of the target file
contains an unexpected value
``admin\textbackslash xd7@example.com'',
which triggers a 500 Internal Server Error (\Cref{res:3}) because
the unicode `x7' is unhandled in the ruby library trying to detokenize and parse
the relative file path. We treat ``500 Internal Server Errors'' as bugs.
To generate new similar test cases with unexpected values, one has to decide
which requests of a test case to mutate, what parts of that request to mutate, and
what new values to inject in those parts.

\newbox\testcaseSample
%\centering
\begin{lrbox}{\testcaseSample}
\begin{lstlisting}[linewidth=0.85\linewidth,language=networklog,basicstyle=\scriptsize, numbers=left,
    stepnumber=1,escapechar=|]
POST /api/projects HTTP/1.1 |\label{req:1}|
Content-Type: application/json
PRIVATE-TOKEN: DRiX47nuEP2AR
{"name":"21a8fa"}

HTTP/1.1 201 Created |\label{res:1}|
{"id":1243, "name":"21a8fa", created_at":"2019-11-23T20:57:15",
"creator_id":1, "forks_count":0, "visibility":"private",
"owner":{"state":"active"}}

POST /api/projects/1243/repository/branches HTTP/1.1 |\label{req:2}|
Content-Type: application/json
PRIVATE-TOKEN: DRiX47nuEP2AR
{"branch":"feature1"}

HTTP/1.1 201 Created |\label{res:2}|
{"branch":"feature1", "commit":{"id":"33c42b", "parent_ids":[],
"title":"Add README.md", "message":"Add README.md",
"author_name":"admin", "authored_date":"2019-11-23T20:57:18"}

POST /api/projects/1243/repository/commits HTTP/1.1 |\label{req:3}|
Content-Type: application/json
PRIVATE-TOKEN: DRiX47nuEP2AR
{"branch":"feature1", "commit_message":"testString",
"actions":[{"action":"create", "file_path":"admin\xd7@example.com"}]}|\label{req3:body}|

HTTP/1.1 500 Internal Server Error  |\label{res:3}|
{"message":"internal server error"}
\end{lstlisting}
\end{lrbox}

\begin{figure}[t]

    \center
    \usebox\testcaseSample
    \caption{
        {\bf \pythia test case and bug found.}
        The test case is a sequence of three API requests
        testing commit operations on \gitlab. After
        creating a new project (first request) and a new branch (second
        request), issuing a commit with an invalid file path triggers an
        unhandled exception.
    }
    \label{fig:testcase}
    \vspace{-10pt}
\end{figure}

\heading{Complexity of REST API testing.}
The example of \Cref{fig:testcase} shows the sequence of events that
need to take place before
uncovering an error. It highlights the complexity of REST API testing
due to the highly-structured, typed format of each API
request and because of producer-consumer dependencies between API
requests. For example, the second request in \Cref{fig:testcase} must include a
structured body payload and also properly use the project id ``1243''
created by the first request. Similarly, the third request must include a
body payload and properly use resources produced by the two
preceding requests (one in its path and one in its body). Syntactic and
semantic validity must be preserved within and across requests of a REST
API test case. Each test case is a {\em stateful} sequence
of requests, since resources produced by preceding requests may be used by
subsequent requests.

\begin{figure}[t]
    \center
    \includegraphics[width=\columnwidth]{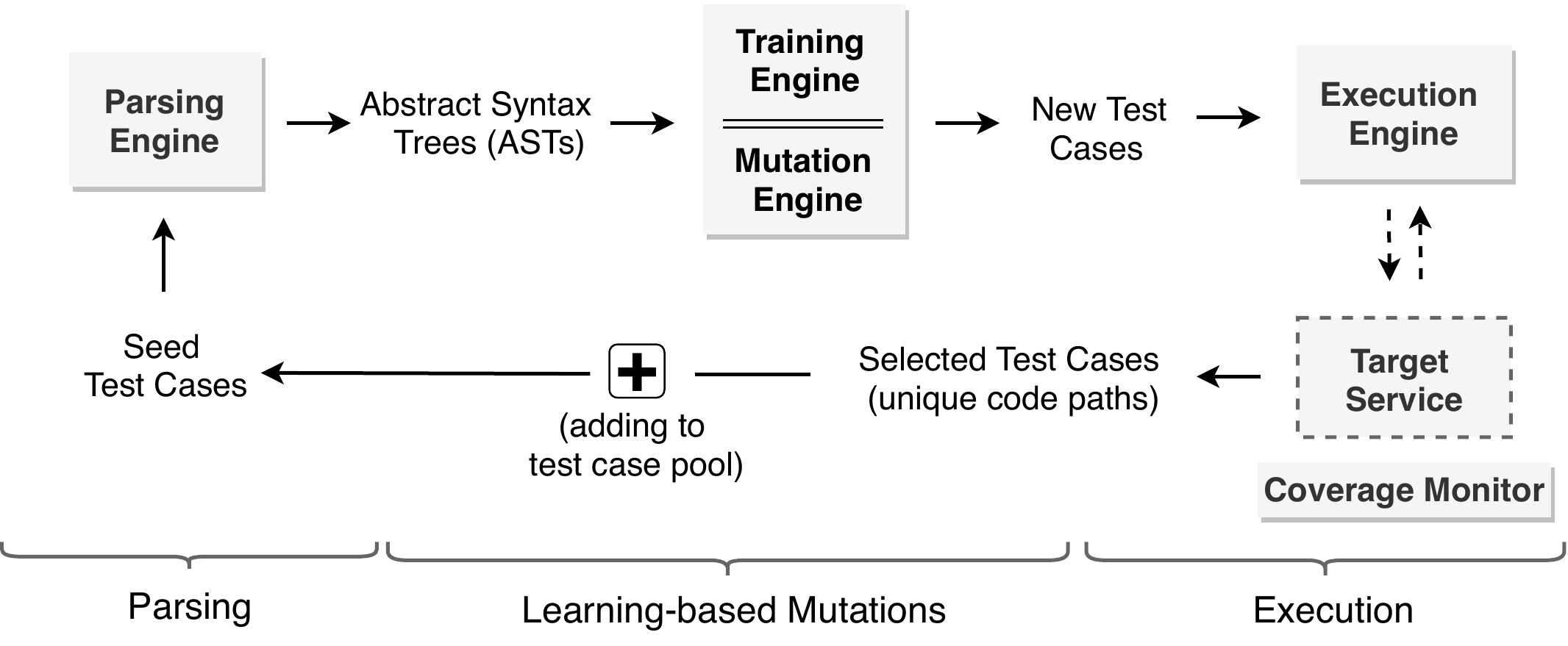}
    \caption{{\bf \pythia architecture.}}
    \label{fig:overview}
    \vspace{-10pt}
\end{figure}

\heading{Existing stateful REST API fuzzing.}
%\heading{Limitations of Existing Approaches.}
Stateful REST API fuzzing, introduced by \restler~\cite{restler}, is a %new
grammar-based fuzzing approach that statically analyzes
the documentation of a REST API (given in an API specification language,
such as OpenAPI~\cite{swagger}), and generates a
{\em fuzzing grammar} for testing a target service through its REST API.
A \restler fuzzing grammar
%is automatically generated {\tt python} source code which
contains rules describing (i) how to fuzz each individual
API request; (ii) what the dependencies are across API requests
and how can they be combined in order to produce longer and longer test cases;
and (iii) how to parse each response and retrieve ids of resources
created by preceding requests in order to make them available to subsequent
requests.
During fuzzing, each request is executed with various value
combinations depending on its primitive types, and the values
available for each primitive type are specified in a user-provided fuzzing dictionary.
%Requests that receive successful responses
%are further combined in order to produce longer and longer test cases.
In the example of \Cref{fig:testcase}, the value of the field ``action'' in the
last request (\Cref{req3:body}) will be one of
``create'', ``delete'', ``move'', ``update'', and ``chmod'' (i.e., the available mutations
for this enum type) and the value of the field ``commit\_message''
will be one of ``testString'' or ``nil'' (the default available mutations for string types).
By contrast, the value of the field ``branch,'' which is a producer-consumer
dependency, will always have the value ``feature1'' created by the previous request.
Thus, the set of grammar rules driving stateful REST API fuzzing leads to
syntactically and semantically valid mutations.

However, \restler, and more broadly this type of grammar-based
fuzzing, has two main limitations.  First, the available mutation
values per primitive type are limited to a small number in order to
limit an inevitable combinatorial explosion in the number of possible
fuzzing rules and values. Second, these static values remain constant
over time and are not prioritized in any way.

\heading{Our contribution.}
To address the above limitations, in the next section, we introduce \pythia,
a new fuzzer that augments grammar-based fuzzing with coverage-guided feedback
and a learning-based mutation strategy for stateful REST API fuzzing.
\pythia's mutation fuzzing strategy generates many new grammatically-valid test
cases, while coverage-guided feedback is used to prioritize test cases that are
more likely to find new bugs.

\section{Pythia}
\label{sec:pythia}
%
% \pythia consists of five main components, namely an
% {\it parsing engine} (\S\ref{sec:parsing-phase}),
% a {\it training engine} (\S\ref{sec:training-engine}),
% a {\it mutation engine} (\S\ref{sec:mutation-engine},)
% an {\it execution engine}
% %(\S\ref{sec:execution-engine}),
% and a {\it coverage monitor} (\S\ref{sec:execution-phase}).

\begin{figure}[t]
\begin{center}
\lstset{%
    %basicstyle=\ttfamily\small\bfseries,
    basicstyle=\ttfamily\scriptsize\bfseries,
    frame=tb,
    mathescape=true
}
\begin{lstlisting}
 $S = sequence$
 $\Sigma = \Sigma_{http-methods} \cup ~\Sigma_{resource-ids} \cup ~\Sigma_{enum}$
  $\cup ~\Sigma_{bool} \cup ~\Sigma_{string} \cup ~\Sigma_{int} \cup ~\Sigma_{static} $
 $N = \{request,~ method,~ path,~ header,~ body,~ \beta_1,~\beta_2,~ \beta_3,$
  $~ producer,~ consumer, ~ fuzzable,~ enum,$
  $~ bool,~ string,~ int, ~ static\}$
 $R = \{sequence \rightarrow  request + sequence ~ | ~ \varepsilon $,
  $ request \rightarrow method + path + header + body$,
  $ method \rightarrow \Sigma_{http-methods}$ , $path \rightarrow  \beta_1 + path ~ | ~ \varepsilon$,
  $ header \rightarrow  \beta_1 + header ~ | ~ \varepsilon$, $ body \rightarrow  \beta_1 + body ~ | ~ \varepsilon$,
  $ \beta_1 \rightarrow  \beta_2 ~ | ~ \beta_3$, $ \beta_2 \rightarrow  producer ~ | ~ consumer$,
  $ producer \rightarrow  \Sigma_{resource-ids}$, $ consumer \rightarrow  \Sigma_{resource-ids}$,
  $ \beta_3 \rightarrow  static ~ | ~ fuzzable$, $ static \rightarrow \Sigma_{static} $,
  $ fuzzable \rightarrow string ~ | ~ int ~ | ~ bool ~ | ~ enum ~ | ~ uuid  $,
  $ string \rightarrow \Sigma_{string}, \dots \}$
\end{lstlisting}
\end{center}
\vspace{-10pt}
\caption{{\small{\bf Regular Grammar (RG) with tail recursion
    for REST API test case generation.}
    The production rules of $\mathcal{G}$ with non-terminal symbols capture the
    properties of any REST API specification,
    while the alphabet of terminal symbols is
    API-specific since different APIs may contain different values for strings,
    integers, enums, and so on.
}}
\vspace{5pt}
\label{fig:rest-api-grammar}
\end{figure}
\begin{figure}
    \centering
    \includegraphics[width=0.49\textwidth]{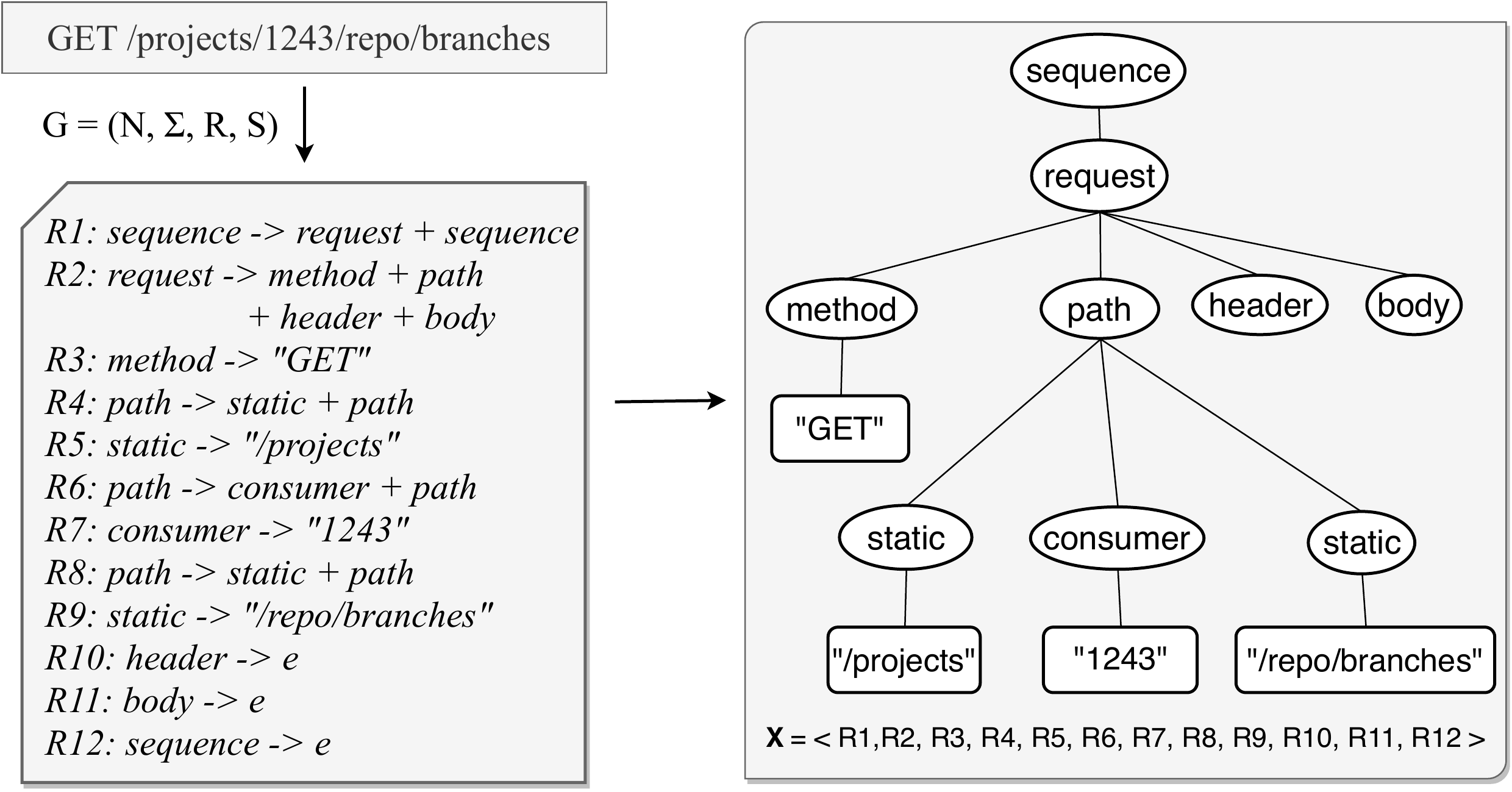}
    \vspace{-10pt}
    \caption{\restler seed test case \& \pythia
        parse tree following $\mathcal{G}$
    }
    \label{fig:abstraction-engine-example}
    \vspace{-5pt}
\end{figure}

%\subsection{Overview}
\pythia is a grammar-based fuzzing engine to fuzz cloud services through
their REST APIs.
Since these APIs are highly structured (see~\Cref{sec:motivation}),
generating meaningful test cases (a.k.a.~mutants) is a non-trivial task\textemdash
the mutants should be structurally valid to bypass initial syntactic checks, yet
must contain some erroneous inputs to trigger unhandled HTTP errors.
Randomly mutating seed inputs often results in invalid structures, as we will
see in~\Cref{sec:evaluation}. One potential solution could be to sample the
mutants from the large space of structurally valid inputs and inject errors to
them. However, for complex grammars, like those defined for REST APIs, exhaustively
enumerating all the valid structures is infeasible. As a workaround, \pythia first
uses a statistical model to learn the common usage of a REST API from seed inputs,
which are all structurally valid. It then injects a small amount of random noise to
deviate from common usage patterns while still maintaining syntactic validity.

\Cref{fig:overview} presents a high-level overview of \pythia.
It operates in three phases: parsing, learning-based mutation, and execution.
First, the parsing phase (\Cref{sec:parsing-phase}) parses the input test
cases using a regular grammar and outputs the corresponding abstract syntax
trees (ASTs). Input test cases can be generated either by using \restler
to fuzz the target service or by using actual production traffic of the target
service.
The next phase, learning-based mutation
(\Cref{sec:learning-phase}), operates on these ASTs. Here, \pythia trains a
sequence-to-sequence (seq2seq) autoencoder~\cite{seq2seq1, seq2seq2}
in order to learn the common structure of the seed test cases. This includes
the structure of API requests (\ie primitive types and values) and the dependencies across requests for a given test case. The mutation engine then mutates the seed test cases such that the mutations
deviate from the common usage, yet obey the structural dependencies.  The mutated test cases are then executed by the {execution engine}. A {coverage monitor} tracks the test case executions on the target
service and measures code coverage. \pythia uses the coverage feedback to select the test cases with unique code paths for further mutations.

\newbox\mutationSampleExternalRule
\begin{lrbox}{\mutationSampleExternalRule}
\begin{lstlisting}[linewidth=0.75\linewidth,
basicstyle=\scriptsize, numbers=left, stepnumber=1,escapechar=|]
Seed test case A
----------
|\textcolor{blue}{\bf POST}| /api/v4/projects/1502252/repository/branches HTTP/1.1
Content-Type: application/json
PRIVATE-TOKEN: DRiX47nuEP2ARa4APFrf
{"ref":"master",|\textcolor{blue}{\bf "branch"}|:"anotherString"}

Mutated test case A'
------------
|\textcolor{red}{\bf GKT}| /api/v4/projects/1502252/repository/branches HTTP/1.1
Content-Type: application/json
PRIVATE-TOKEN: DRiX47nuEP2ARa4APFrf
{"ref":"master",|\textcolor{red}{\bf "devexf1opers\_can\_merge"}|:"anotherString"}
\end{lstlisting}
\end{lrbox}

\newbox\mutationSampleLocalRule
\begin{lrbox}{\mutationSampleLocalRule}
\begin{lstlisting}[linewidth=0.7\linewidth,
basicstyle=\scriptsize, numbers=left, stepnumber=1,escapechar=|]
Seed test case B
----------
POST /spree\_oauth/token HTTP/1.1
Content-Type: application/json
{"username":"spree@ex.com","grant_type":|\textcolor{blue}{\bf "password"}|,
"password":"spree123"}

Mutated test case B'
------------
POST /spree\_oauth/token HTTP/1.1
Content-Type: application/json
{"username":"spree@ex.com",
"grant_type":|\textcolor{red}{\bf "xf8pree@ex.com"}|, "password":"spree123"}
\end{lstlisting}
\end{lrbox}

%\begin{figure}[t]
%    \centering
%    \usebox\mutationSampleExternalRule
%    \vspace{-7pt}
%    \caption{{\bf
%    %Sample seed test case and example mutations with new values
%    Example of mutations with new values
%    that are *not*  available in the original test cases.
%    The first mutation changes the request type from POST to GET and
%    further pollutes it with random bytes. This leads an unhandled HTTP
%    request type GTK. The second
%    mutation changes the dictionary key ``branch''
%    using the value ``developers\_can\_merge'' which is a key
%    belonging to a completely different request. The later is further polluted
%    with random bytes that turn it into
%    %the unhandled value
%    ``devexf1opers\_can\_merge''.
%    }}
%    \label{fig:mutationExternal}
%    \vspace{-10pt}
%\end{figure}
%
%\begin{figure}[t]
%    \centering
%    \usebox\mutationSampleLocalRule
%    \vspace{-5pt}
%    \caption{{\bf
%        Example of mutation using values available
%        for the primitive types of the original test case. The grant type value
%        ``password'' is mutated using the username value ``spree@example.com''
%        (available in the same dictionary definition) and is further polluted
%        with random bytes that eventually turn it
%        into ``xf8pree@example.com''
%    }}
%    \label{fig:mutationInternal}
%    \vspace{-15pt}
%\end{figure}

\begin{figure}[t]
    \center
    \usebox\mutationSampleExternalRule
    \caption{{\bf
    %Sample seed test case and example mutations with new values
    Mutations with new values
    that are *not* in the original test cases.
    The first mutation changes the request type from POST to GET and
    further pollutes it with random bytes. This leads an unhandled HTTP
    request type GTK. The second
    mutation changes ``branch''
    using the value ``developers\_can\_merge'' from
    a completely different request definition. The later is further polluted
    with random bytes that turn it into
    %the unhandled value
    ``devexf1opers\_can\_merge''.
    }}
    \label{fig:mutationExternal}
    \vspace{-10pt}
\end{figure}
%\subsection{API Abstraction Engine}
%\subsection{API Abstraction Phase}
\subsection{Parsing Phase}
\label{sec:parsing-phase}
In this phase, \pythia infers the syntax of the seed inputs by parsing them
with a user-provided
%regular grammar.
%We model the generation of REST API test cases using a
Regular Grammar
(RG) with tail recursion. Such an RG is defined by a $4$-tuple
$\mathcal{G}= (N, \Sigma, R, S)$, where $N$ is a set of non-terminal symbols,
$\Sigma$ is a set of terminal symbols, $R$ is a finite set of production rules
%$of the form $A \rightarrow  A~|~aB~|~\epsilon$,
%$where $A,B \in N$, $\alpha \in \Sigma$, and  $\epsilon$ denotes the empty string.
of the form $\alpha \rightarrow \beta_1\beta_2\dots\beta_n$,
where $\alpha \in N, n \ge 1$, $\beta_i \in (N \cup \Sigma), \forall 1 \le i \le n$,
and $S \in  N$ is a distinguished start symbol. The syntactic definition
of $\mathcal{G}$ looks like a Context Free Grammar, but because recursion is
only allowed on the right-most non-terminal and there are no other cycles
allowed, the grammar is actually regular.
% \begin{itemize}%[leftmargin=*]
%   \item $N$ is a set of non-terminal symbols,
%   \item $\Sigma$ is a set of terminal symbols,
%   \item $R$ is a finite set of production rules of the form
%         $\alpha \rightarrow \beta_1\beta_2\dots\beta_n$, where $\alpha \in N, n \ge 1$,
%         and $\beta_i \in (N \cup \Sigma), \forall 1 \le i \le n$,
%   \item and $S \in  N$ is a distinguished start symbol.
% \end{itemize}
\Cref{fig:rest-api-grammar} shows a template $\mathcal{G}$ for
REST API test
case generation. A test case that belongs to the language defined by
$\mathcal{G}$ is a sequence starting with the symbol {\em sequence} followed by
a successions of production rules ($R$) with non-terminal symbols ($N$) and
terminal symbols ($\Sigma$).

%\heading{Tree representation of REST API test cases.}
\Cref{fig:abstraction-engine-example} shows how seed \restler test cases are parsed by \pythia's parsing engine. The successions of production rules in $\mathcal{G}$ (see LHS of \Cref{fig:abstraction-engine-example}) are applied to infer the corresponding Abstract Syntax Trees (ASTs) (see RHS); the tree  internal nodes are nonterminals, and the leaves are terminals of $\mathcal{G}$.  \pythia parses the tree in Depth First Search (DFS) order, which represents a sequence of grammar rules. For example, a simple test case {\tt X=``GET /projects/1243/\-repo/branches"} will be represented as a sequence of grammar production rules $\mathcal{X}=< R_1, R_2, \ldots, R_{12} >$, as shown in the Figure.
Given a set of seed inputs, thus, the output of this phase is a set of abstracted test cases,
$\mathcal{D}=\{\mathcal{X}_1, \mathcal{X}_2, \dots, \mathcal{X}_{N}\}$, which is passed to the training and mutation engines.

\begin{figure}[t]
    \center
    \usebox\mutationSampleLocalRule
    \caption{{\bf
        Mutations with values available
        for the primitive types of the original test case. The value
        ``password'' is mutated using the value ``spree@example.com''
        (available in the same seed) and is further polluted
        with random bytes that turn it
        into ``xf8pree@example.com''
    }}
    \label{fig:mutationInternal}
    \vspace{-10pt}
\end{figure}
\subsection{Learning-based Mutation Phase}
\label{sec:learning-phase}

The goal of this phase is first to learn the common structural patterns of the target APIs from the seed inputs (see~\Cref{sec:training-engine}), and then to mutate those structures (see~\Cref{sec:mutation-engine}) and generate new test cases. To learn the structural patterns from the existing test cases, \pythia uses an autoencoder model, $\mathcal{M}$, which is trained with the  ASTs of the seed inputs ($\mathcal{D}$).
An autoencoder consists of an encoder and a decoder (see~\Cref{fig:mutation-engine-example2}).
$\mathcal{M}_{encoder}$ represents an abstracted test case $\mathcal{X} \in \mathcal{D}$
to an embedded feature space $\mathcal{Z}$, which captures
the latent dependencies of $\mathcal{X}$.
$\mathcal{M}_{decoder}$ decodes $\mathcal{Z}$ back to $\mathcal{X}'$.
To generate structurally valid mutants, \pythia then minimally perturbs the embedded feature $\mathcal{Z}$ and decodes it back to original space, say $\mathcal{X}'$. Our key insight is that since the decoder is trained to learn the grammar, the output of the decoder from the perturbed hidden state will still be syntactically valid.
Thus, $\mathcal{X}'$ will be syntactically valid mutant.
This section illustrates this design in details.

\subsubsection{Training Engine}
\label{sec:training-engine}

Given the abstracted test cases, $\mathcal{D}$, the training engine learns
their vector representations (i.e., encoding) using an autoencoder type of
neural network~\cite{hinton2006autoencoders}. \pythia realizes the autoencoder
with a simple seq2seq model
$\mathcal{M_\mathcal{D}}$ trained over $\mathcal{D}$. Usually, a seq2seq model
is trained to map variable-length sequences of one domain to another
(\eg English to French). By contrast, we train $\mathcal{M}$ only on
sequences of domain $\mathcal{D}$ such that $\mathcal{M_\mathcal{D}}$
captures the latent characteristics of test cases.

% of one particular REST API.

%\heading{Learning tree representations.}
A typical seq2seq model
%is shown in~\F\ref{fig:seq2seq-example}. It
consists
of two Recurrent Neural Networks (RNNs): an encoder RNN  and a decoder RNN.
% \begin{figure}[h]
%     \center
%     %\includegraphics[width=0.45\textwidth]{figures/seq2seq-example.png}
%     \includegraphics[width=0.45\textwidth,height=0.12\textheight]{figures/seq2seq-example.png}
%     %\vspace{5pt}
%     \caption{{\bf Overview of a typical sequence-to-sequence model.}}
%     \label{fig:seq2seq-example}
%     \vspace{-10pt}
% \end{figure}
The encoder RNN consists of
a hidden state ${\bf h}$ and an optional output ${\bf y}$, and operates
on a variable length input sequence ${\bf x}=<x_1, \dots, x_n>$.
%augmented with two auxiliary tokens {\tt <SOS>} and {\tt <EOS>}
%marking the beginning and the ending of each sequence respectively.
At each time t (which can be thought of as position in the sequence),
the encoder reads sequentially each symbol $x_{t}$
of input ${\bf x}$, updates its hidden state ${\bf h_{t}}$ by ${{\bf h}_{t}}=f({\bf h}_{t-1},~x_t)$,
% \begin{equation}
%     {{\bf h}_{t}}=f({\bf h}_{t-1},~x_t),
%     \label{eq:enc-hidden}
% \end{equation}
where f is a non-linear activation function, such as a simple
%REctified Linear Unit (ReLU) or a more complex
a Long Short-Term Memory
(LSTM) unit~\cite{lstm}, and calculates the output $y_t$ by $y_{t}=\phi({\bf h}_{t})$,
% \begin{equation}
%     y_{t}=\phi({\bf h}_{t}),
% \label{eq:enc-output}
% \end{equation}
where $\phi$ is an activation function producing valid probabilities.
% such as a {\tt softmax}~\cite{softmax}.
% that computes the output probability distribution over
% a given vocabulary conditioned on the current hidden state.
At the end of each input sequence,
the hidden state of the encoder is a summary ${\bf z}$ of the whole
% input
sequence. Conversely, the decoder RNN
generates an output sequence ${\bf y}=<y_1, \dots, y_{n'}>$
by predicting the next symbol $y_{t}$ given the hidden state ${\bf h}_{t}$,
where both $y_{t}$ and ${\bf h}_{t}$
are conditioned on $y_{t-1}$ and on the summary ${\bf z}$ of the input
sequence. Hence, the hidden state of the decoder at time t is computed by ${\bf h}_{t}=f({\bf h}_{t-1},~y_{t-1},{\bf z})$,
% \begin{equation}
%     {\bf h}_{t}=f({\bf h}_{t-1},~y_{t-1},{\bf z}),
% \label{eq:dec-hidden}
%\end{equation}
and the conditional distribution of the next symbol is computed by
%\begin{equation}
    $y_{t}=\phi({\bf h}_{t},~y_{t-1},{\bf z})$
% \label{eq:dec-output}
% \end{equation}
for given activation functions $f$ and $\phi$.
% The complete sequence ${\bf y}$
% is produced by iteratively applying this equation %eq.~\ref{eq:dec-output}
% until {\tt <EOS>} is emitted.

We jointly train a seq2seq model $\mathcal{M}$ on $\mathcal{D}$
to maximize the conditional log-likelihood
% \begin{equation*}
$\argmaxA_{{\boldsymbol \theta}} \frac{1}{N}\sum_{i=1}^{N} log p_{{\boldsymbol \theta}} ({\bf y}_i | {\bf x}_i)$,
% \end{equation*}
where ${\boldsymbol \theta}$ is the set of the learnt
model parameters and each ${\bf x}_i, {\bf y}_i \in \mathcal{D}$.
As explained earlier, $\mathcal{M}_{\theta, \mathcal{D}}$ is
trained on sequences of one domain (\ie ${\bf y}_i={\bf x}_i$) and is then
given as input to the mutation engine.

\subsubsection{Mutation Engine}
\label{sec:mutation-engine}
\newcommand\mycommfont[1]{\footnotesize\textcolor{blue}{#1}}
\SetCommentSty{mycommfont}
\SetKwComment{Comment}{}{}
{
\footnotesize
\setlength{\textfloatsep}{0pt}
\begin{algorithm}[t]
\SetAlgoLined
    \KwIn{seeds $\mathcal{D}$, RG grammar $\mathcal{G}$, model $M_{\theta, \mathcal{D}}$, batch size $N$}

    %\Comment*[h]{// Iterate over all seeds until time budget is  expired}

   \While{time\_budget}
   {{\label{alg:budget}}
        $\mathcal{X} \leftarrow get\_next(\mathcal{D})$

        $\mathcal{Z} \leftarrow M_{\theta, \mathcal{D}}.endoder(\mathcal{X})$

        $new\_sequences \leftarrow \emptyset$

        \Comment*[h]{//Perturbation: Exponential search on random noise scale}

        \For{$j\gets0$ \KwTo $N$}{ {\label{alg:noise_start}}

             \Comment*[h]{// Noise draw from normal distribution}

            $\delta_j \leftarrow random.normal(\mathcal{Z}.shape, 0)$

             \Comment*[h]{// Bound and scale random noise}

            $\delta_j \leftarrow 2^j *  \delta_j / \|\mathcal{Z}\|_{2} $

             \Comment*[h]{// Add noise on decoder's starting state}

            $\mathcal{X}_{j}' \leftarrow M_{\theta, \mathcal{D}}.decoder( \mathcal{Z} + \delta_j)$

            $new\_sequences.append(\mathcal{X}_{j}')$
            % \If{$\hat{\mathcal{X}_{i, j}}~!=~\mathcal{X}$}{
            %     $predictions.append(\hat{\mathcal{X}})$
            % }
        }

         \Comment*[h]{// Select the prediction with smallest noise scale}

        $\mathcal{X}_{min}' \leftarrow \argminA_{scale} new\_sequences$ {\label{alg:noise_end}}

         \Comment*[h]{// Case 1: Grammar rules not seen in the current seed}

        $rules \leftarrow terminals(\mathcal{G}) - terminals(\mathcal{X})$ {\label{alg:mut_start}}

        \ForEach{ index {\bf in} get\_common\_leafs($\mathcal{X}$,~$\mathcal{X}_{min}'$)}{ {\label{alg:muta_start}}
            \ForEach{ rule {\bf in} rules}{

                $mutation \leftarrow rule + random\_bytes$

                $\mathcal{X}[index] \leftarrow mutation$

                $EXECUTE(\mathcal{X})$
            }
        } {\label{alg:muta_end}}

         \Comment*[h]{// Case 2: Grammar rules from the new decoder's prediction}

        $rules \leftarrow terminals(\mathcal{X}_{min}')$

        \ForEach{index {\bf in} get\_different\_leafs($\mathcal{X}$,~$\mathcal{X}_{min}'$)}
        {  {\label{alg:mutb_start}}
            \ForEach{ rule {\bf in} rules}{

                $mutation \leftarrow rule + random\_bytes$

                $ \mathcal{X}[index] \leftarrow mutation$

                $EXECUTE(\mathcal{X})$
            }
        }{\label{alg:mut_end}}
    }
    \caption{{\bf Learning-based \pythia mutations}}
    \label{alg:guided-tree-mutations}
    %\vspace{-5pt}
\end{algorithm}
%\vspace{-5pt}
}

\begin{figure}[t]
    \center
    \includegraphics[width=0.49\textwidth]{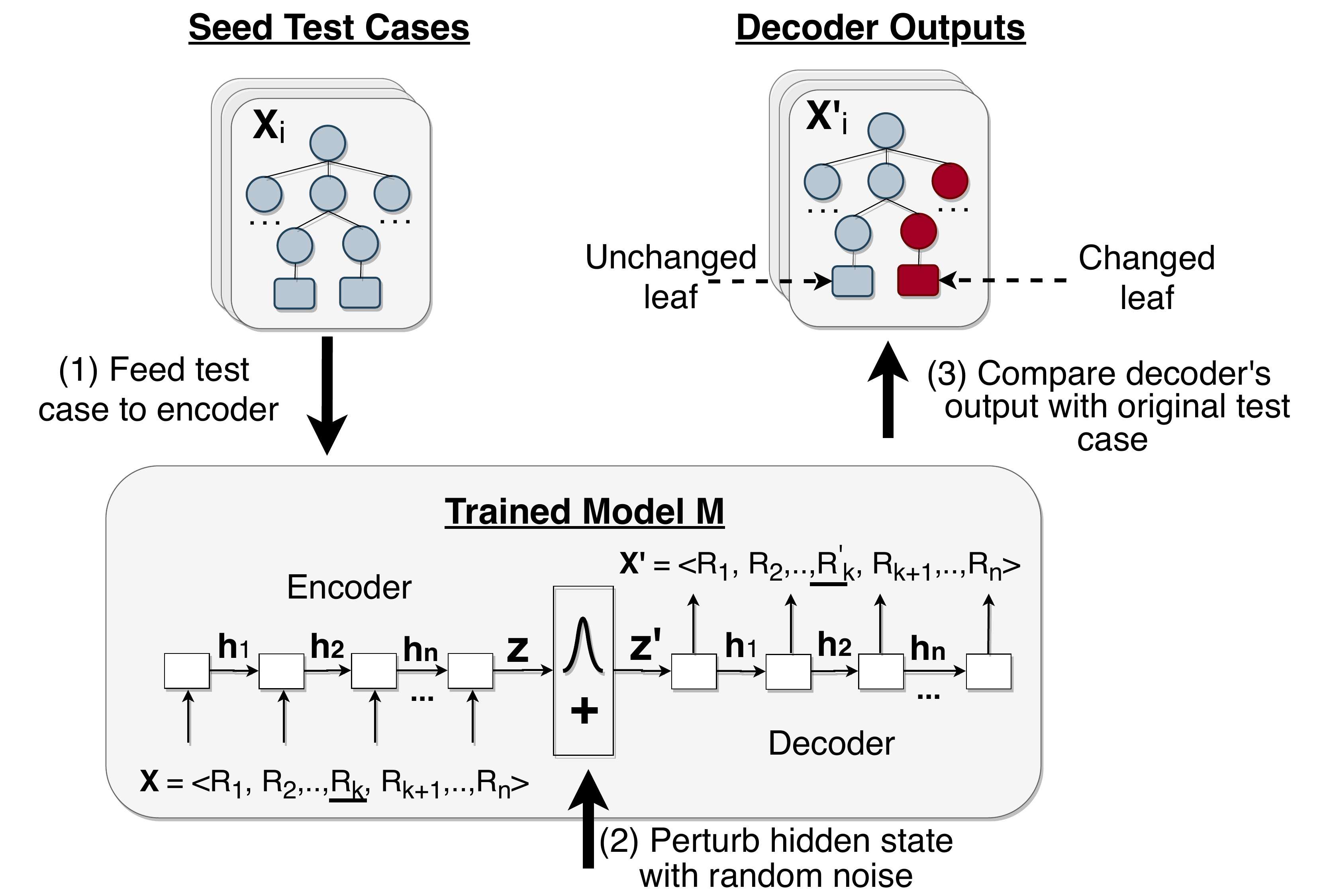}
    \caption{{\bf Overview of Pythia Mutation Engine}}
    \label{fig:mutation-engine-example2}
\end{figure}

For each test case  $\mathcal{X} \in \mathcal{D}$, the mutation engine decides
with what values to mutate each input location of $\mathcal{X}$.
Since $\mathcal{X}$ is a sequence of grammar rules
$<R_1, R_2, \ldots , R_n>$ (see~\Cref{fig:abstraction-engine-example}), the
mutation strategy determines how to mutate each rule:
whether to use alternative rules with different values not present in the
current test case (example of~\Cref{fig:mutationExternal})
or use rules available in the original seed test case
(example of~\Cref{fig:mutationInternal}).
To mutate a seed test $\mathcal{X}$, \pythia first perturbs its embedded vector
representation ($\mathcal{Z}$) by adding minimal random noise,
and decodes it back to a new test case $\mathcal{X}'$. The perturbation added
by \pythia  may create differences between $\mathcal{X}$ and $\mathcal{X}'$.
These differences determine the mutation strategy on each location of the
seed test case:
\begin{itemize}
    \setlength{\itemindent}{0.25in}
    \item[\underline{Case 1:}]
    Locations where $\mathcal{X}'$ and $\mathcal{X}$ are the same after
    perturbation indicate that the model has not seen many variations during
    training and mutations with new rules, not in the original input/output
    sequences, should be used (See example of \Cref{fig:mutationExternal}).
    \item[\underline{Case 2:}]
    Locations where $\mathcal{X}'$ and $\mathcal{X}$ differ
    indicates that the model has seen more variance during training and
    mutations with the rules seen by the model should be used
    (See example of \Cref{fig:mutationInternal}).
    In fact, these rules are auto-selected from the decoder.
    %during the decoder phase.
\end{itemize}

Algorithm~\ref{alg:guided-tree-mutations} presents the mutation strategy in details and \F\ref{fig:mutation-engine-example2} pictorially illustrates it.  The algorithm takes a set of abstracted
test cases $\mathcal{D}$, a regular grammar $\mathcal{G}$, a trained autoencoder model $\mathcal{M}_{\theta, \mathcal{D}}$ and its batch size $N$ as inputs, and continuously iterates over $\mathcal{D}$ until the time budget expires (Line~\ref{alg:budget}).
At a high level, the mutation engine has two steps: identifying mutation types
appropriate for each location and applying the changes to those locations.

\noindent

$\bullet$ \textit{Perturbation } (lines~\ref{alg:noise_start} to
\ref{alg:noise_end}):
%of Algorithm\ref{alg:guided-tree-mutations}):
For each test case $\mathcal{X}$, the encoder of model $M_{\theta, \mathcal{D}}$
obtains its embedding $\mathcal{Z}$ and then the embedded vector is perturbed
with random noise.  In particular, \pythia draws $N$ noise-values
$\{\delta_0, \delta_1, \dots, \delta_{N-1}\}$ from a normal distribution,
bounded by $2-$norm of $\mathcal{Z}$ and scaled exponentially
in the range $\{2^0, 2^1, \dots, 2^{N-1}\}$. The $N$ noise values
are used to perturb $\mathcal{Z}$ independently $N$ times and get different
perturbed vectors $\{\mathcal{Z} + \delta_0,~ \mathcal{Z} + \delta_1, \dots,~\mathcal{Z} + \delta_{N-1}\}$,
which serve as $N$ different starting states of the decoder.
In turn, they lead to $N$ different output sequences
$\{\mathcal{X}_{0}',~ \mathcal{X}_{i}', \dots, ~\mathcal{X}_{N-1}'\}$
for each input $\mathcal{X}$. From these $N$ new outputs,
\pythia selects $\mathcal{X}_{min}'$ which differs from
$\mathcal{X}$ and is obtained by the smallest ($2$-norm) perturbation
$\delta_{min}$ on $\mathcal{Z}$.

The $N$-step exponential search performed in order to find the smallest
perturbation that leads to a new prediction $\mathcal{X}_{min}'$ helps avoid
very pervasive changes that will completely destroy the embedded
structure $\mathcal{Z}$ of $\mathcal{X}$. Generally, norm-bounded perturbations
is a common approach in the literature of adversarial machine learning
~\cite{biggio2013evasion,goodfellow2014explaining,carlini2017towards} where,
given a classification model $f$ and an input sample $x$ originally classified
to class $f(x)$, the goal is to find a small perturbation $\delta$ that will
change the original class of $x$ such that $f(x+\delta) \ne f(x)$. Our use of
perturbations in Algorithm~\ref{alg:guided-tree-mutations} is different in two
ways. First, the perturbations are random as opposed to typical adversarial
perturbations that are guided by the gradients of the classification model $f$.
Second, the seq2seq model $M_{\theta, \mathcal{D}}$ is not a classification
model, but rather an autoencoder. The purpose of applying perturbations on the
initial state of the decoder is, given a seed test case $\mathcal{X}$,
to leverage the
knowledge learnt from $\mathcal{M}_{\theta, \mathcal{D}}$ on $\mathcal{D}$
and generate a new test case $\mathcal{X}_{min}'$ that is marginally different
from the original one. We then compare $\mathcal{X}_{min}'$ and
$\mathcal{X}$ to determine the mutation strategy on each location of the seed
test case.

$\bullet$ \textit{Comparison \& Mutation Strategies}
(lines~\ref{alg:mut_start} to \ref{alg:mut_end}):
%Algorithm~\ref{alg:guided-tree-mutations}):
The result of the comparison between $\mathcal{X}_{min}'$ and $\mathcal{X}$
determines the mutation strategy followed on each location of the seed
test case. The two groups of nested for-loops implement the two
different mutation strategies explained earlier.
The first group of nested for-loops in the Algorithm targets leaf
locations where $\mathcal{X}_{min}'$ and
$\mathcal{X}$  are the same (Case 1). For such positions
of $\mathcal{X}$ (lines~\ref{alg:muta_start} to \ref{alg:muta_end}), new
mutations are generated by iteratively applying grammar rules with terminal
symbols originally not in $\mathcal{X}$.
The second group of nested for-loops (lines~\ref{alg:mutb_start} to
\ref{alg:mut_end}) targets leaf locations where $\mathcal{X}$ and
$\mathcal{X}_{min}'$ differ (Case 2). For such positions of $\mathcal{X}$ new mutations
are generated by iteratively applying grammar rules with terminal symbols
in $\mathcal{X}_{min}'$. In both cases, the new grammar rules are further
augmented with random byte alternations on the byte representation of each
rule's terminal symbols. This augmentation with
auxiliary payload mutations helps avoid repeatedly exercising identical rules.

\subsection{Execution Phase}
\label{sec:execution-phase}

%\subsubsection{Execution Engine}
%\label{sec:execution-engine}

In this step the execution engine takes as inputs new test cases generated by
the mutations engine and executes them to the target service which is
continuously being monitored by the coverage monitor. Executing a test case
includes sending its requests to the target service over http(s) and receiving
back the respective responses. Before testing, we statically analyze the
source code of the target service, extract basic block locations, and configure
it to produce code coverage information. During testing, the coverage monitor
collects code coverage information produced by the target service and matches it
with the respective test cases executed by the execution engine. Then, given the
basic blocks statically extracted, each test case is mapped into a bitmap of
basic blocks describing the respective code path activated. This helps
distinguish test cases that reach new code paths and
ultimatelly minimize an initially large corpus of seed test cases to a smaller
one with test cases that reach unique code paths.

\subsection{Implementation}
We use an off-the-shelf seq2seq RNN
with input embedding, implemented in tensorflow~\cite{tensorflow}.
The model has one layer of  $256$ Gated Recurrent Unit (GRU) cell in the encoder
as well as in the decoder. Dynamic input unrolling is performed using
{\tt tf.nn.dynamic} RNN APIs and the encoder is initialized with a zero state.
We train the model by minimizing the weighted cross-entropy loss for sequences
of logits using the Adam optimizer~\cite{adam}. We use batches of $32$
sequences, iterate for $2000$ training steps with a learning rate of $0.001$,
and an initial embedding layer of size $100$. The vocabulary of the model
depends on the number of production rules in the fuzzing grammar of each API
family and ranges in couple of hundred of production rules. Similarly, the
length of each sequence depends on the specific API and ranges from
$505$ to $825$ items.
%
% [VA]: specify the APIs.
%
Training such a model configuration
in a CPU-only machine takes no more than two hours.
All the experiments discussed in our evaluations were run on Ubuntu 18.04
Google Cloud VMs~\cite{google-cloud} with 8 logical CPU cores
and 52GB of physical memory.
%, without any Graphics Processing Unit (GPU) acceleration.
Each fuzzing client is used to test a target
service deployment running on the same machine.

\begin{figure*}[!htpt]
    \begin{minipage}{\textwidth}
    \centering
    \subfigure{
        \includegraphics[width=0.3\textwidth]{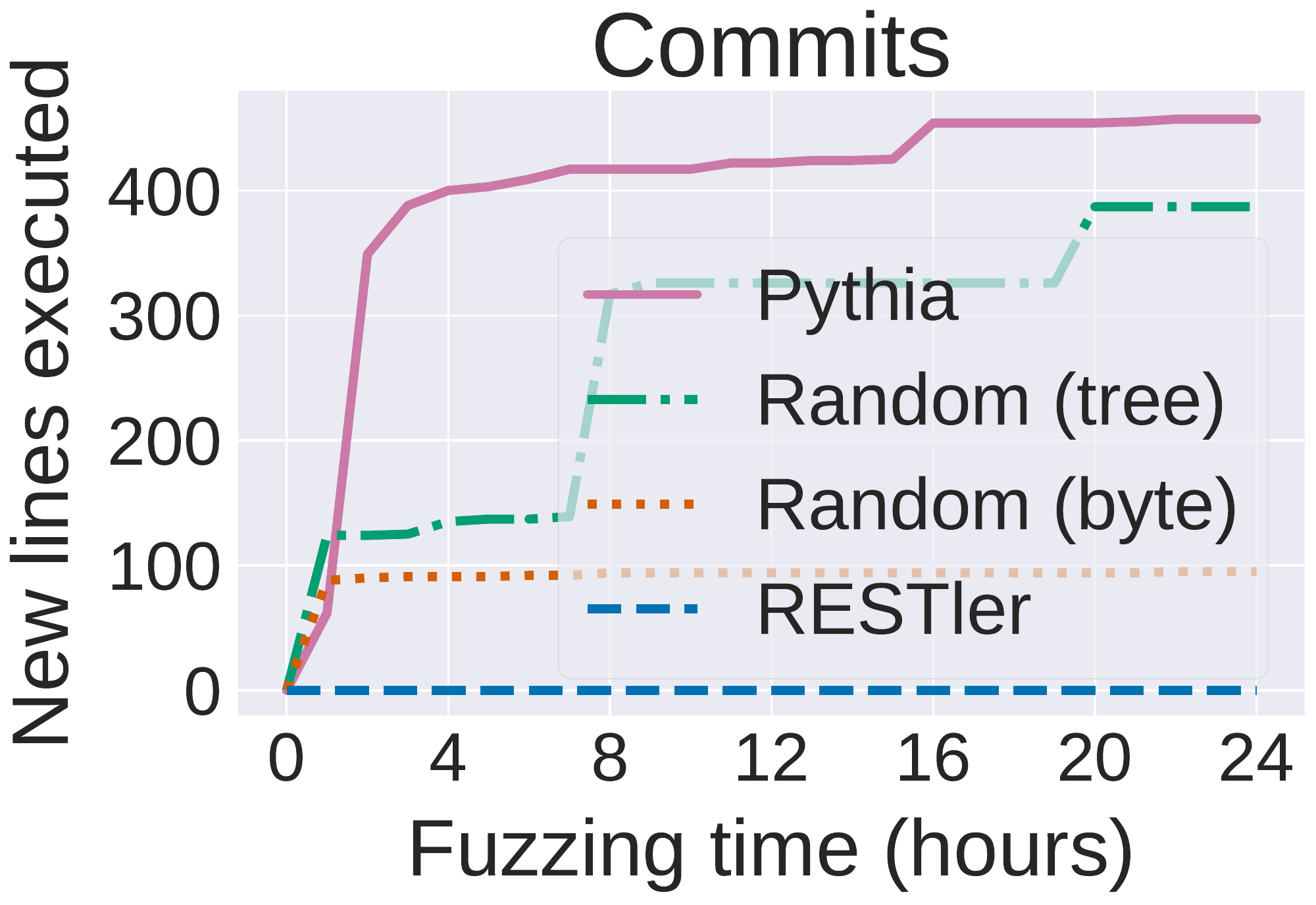}
    }
    \subfigure{
        \includegraphics[width=0.3\textwidth]{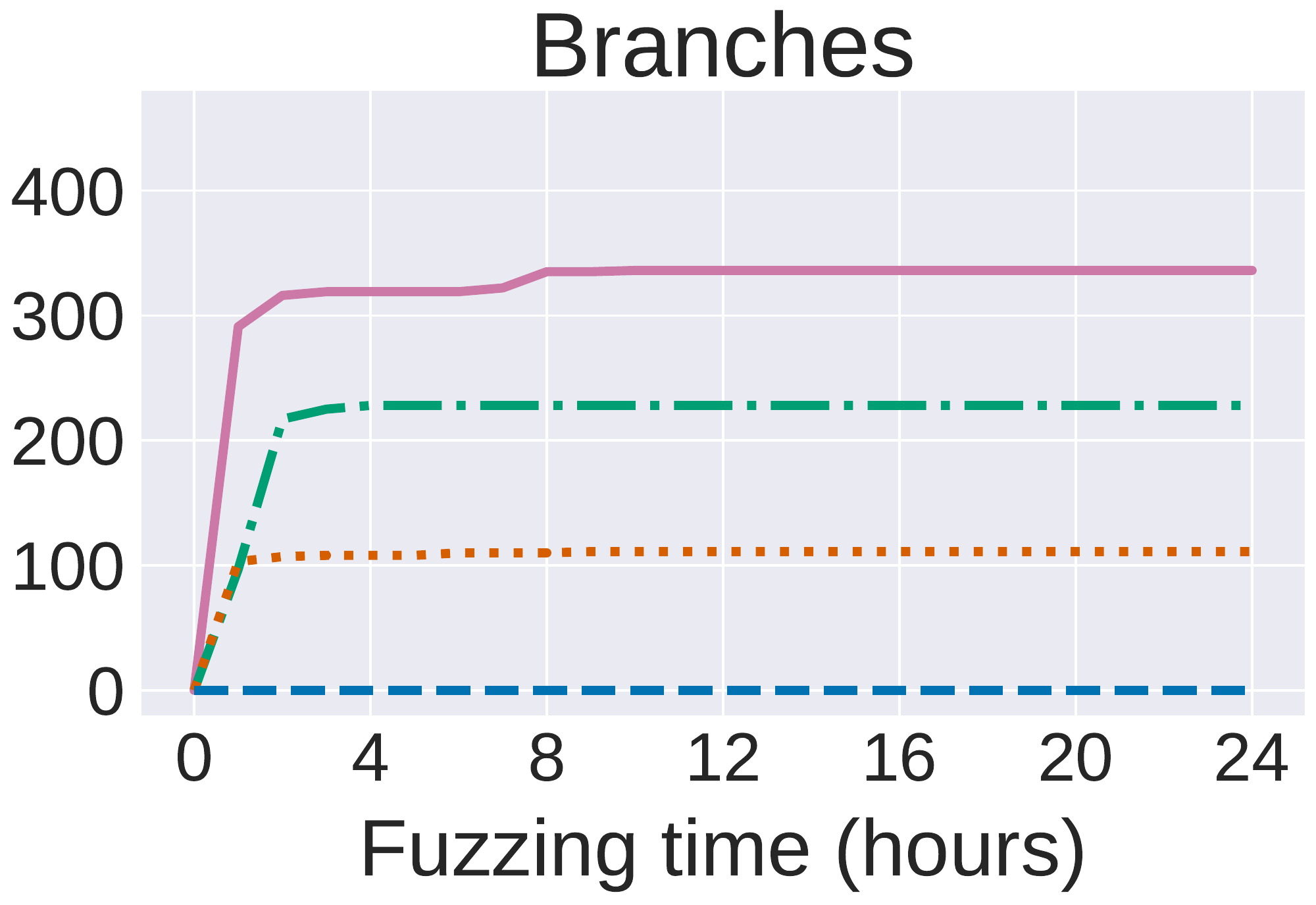}
    }
    \subfigure{
        \includegraphics[width=0.3\textwidth]{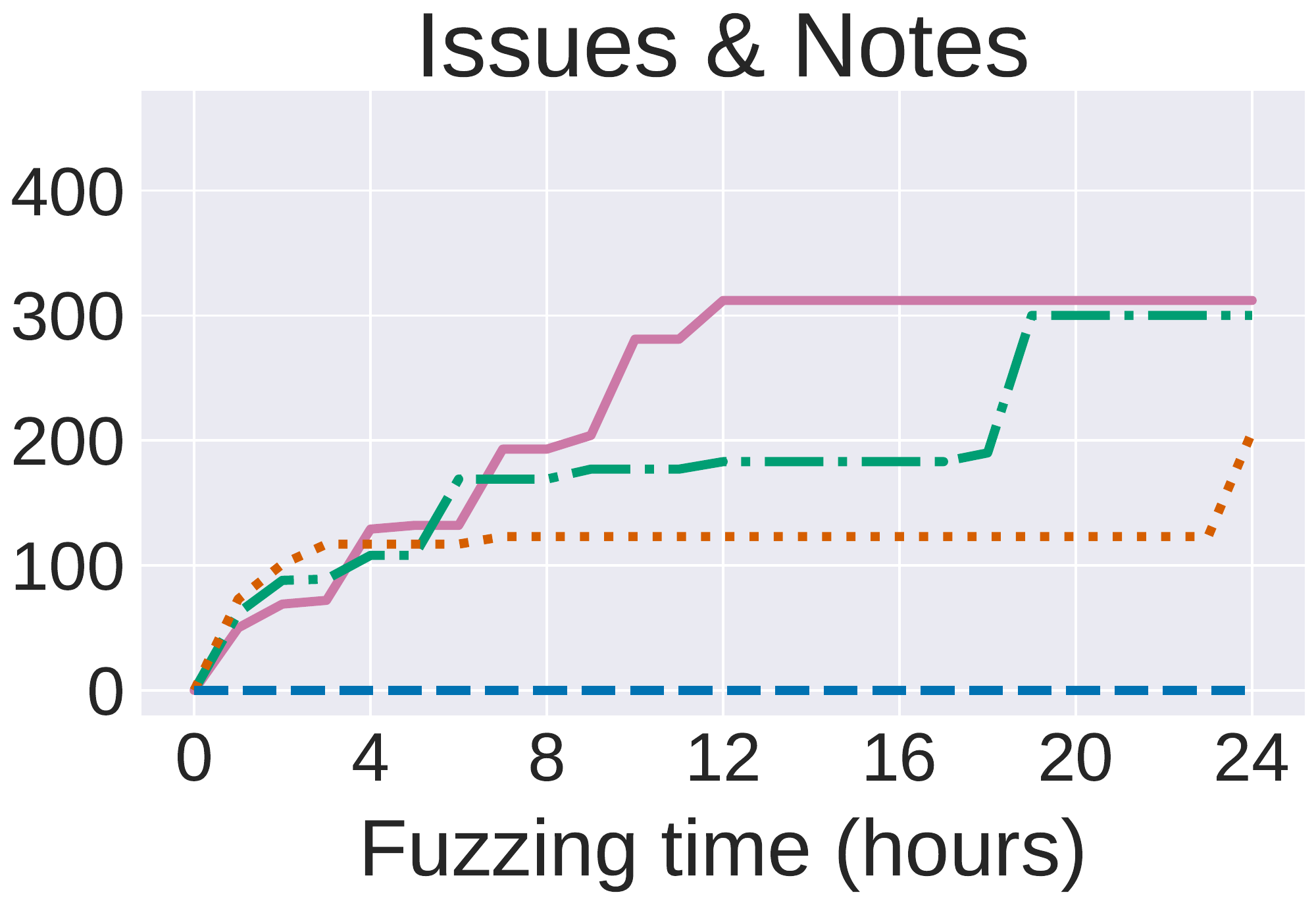}
    }

    \medskip
    \subfigure{
        \includegraphics[width=0.3\textwidth]{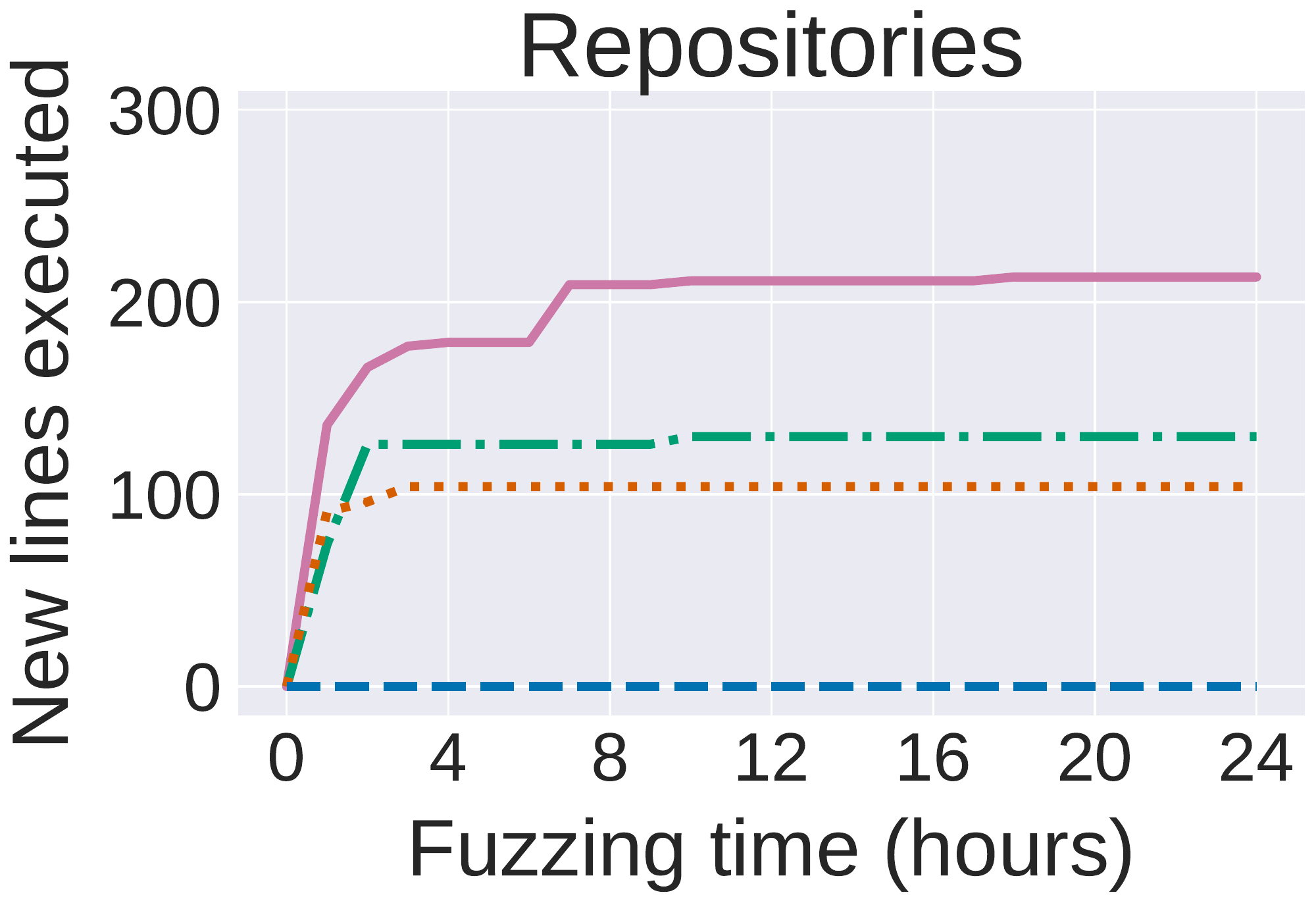}
    }
    \subfigure{
        \includegraphics[width=0.3\textwidth]{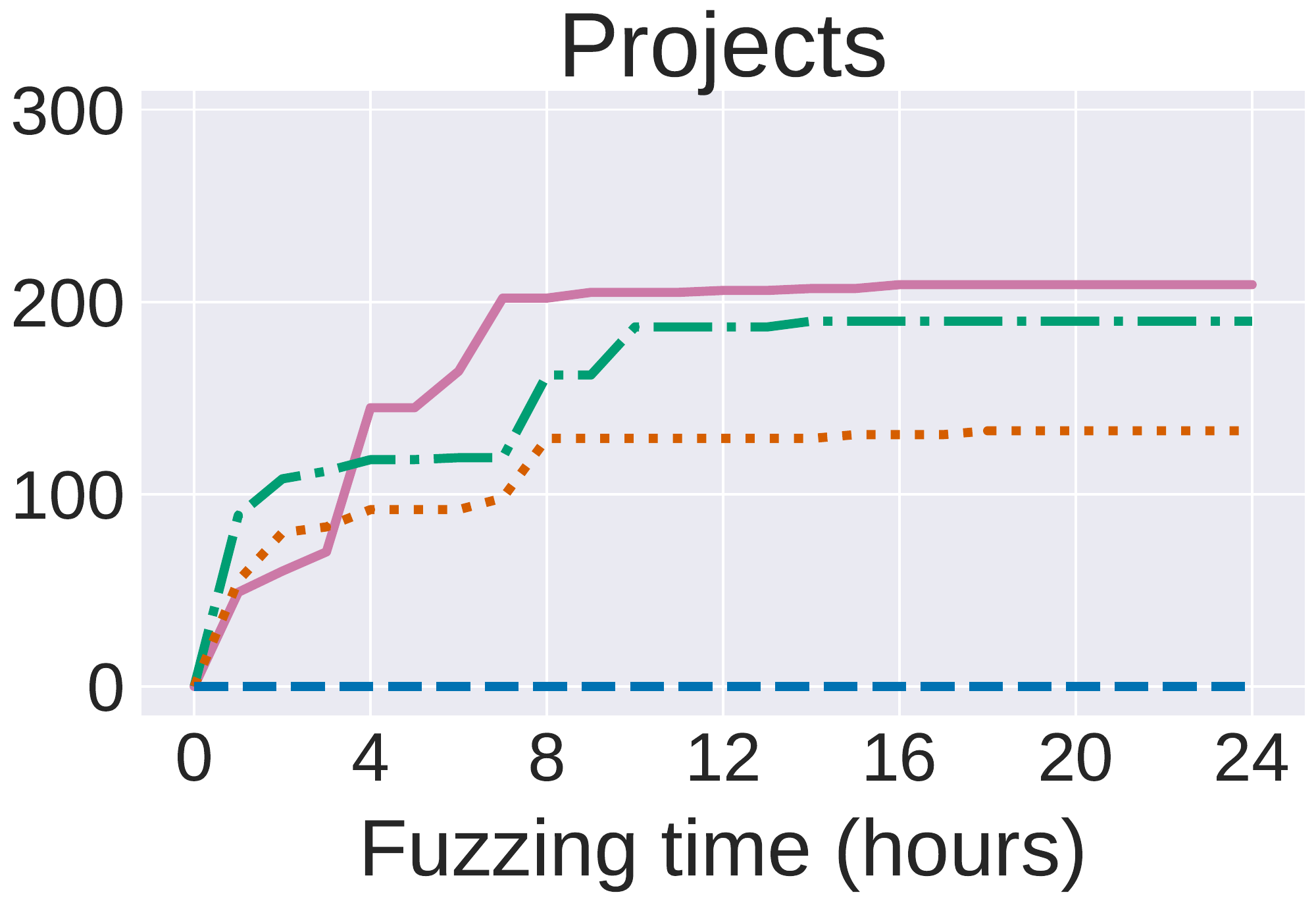}
    }
    \subfigure{
        \includegraphics[width=0.3\textwidth]{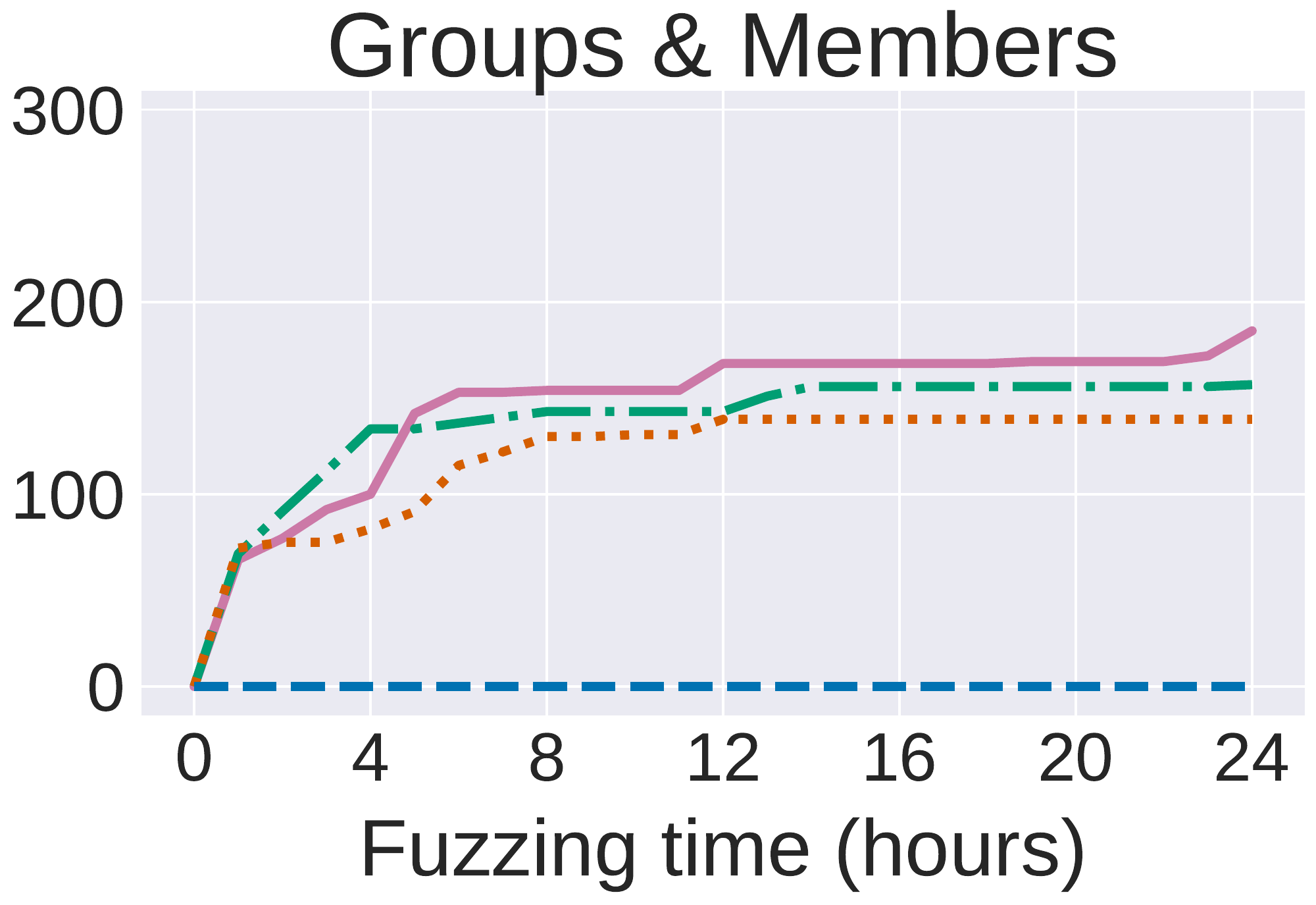}
    }
    \caption{
        {\bf RQ1.~Comparison of \pythia mutations strategies w.r.t.~other baselines.}
        \underline{Seed collection:} Run \restler on each API to generate seed test cases. The seed collection time is set to $24$h for all APIs except for ``issues" in which the respective time was extended to $32$h. Within this time, \restler reached a plateau for all the cases.
\underline{Fuzzing:} Use seed corpus to perform three individual         $24$h fuzzing sessions per API and let \restler also run for an additional $24$h additional hours.
\underline{Comparison:} Measure the number of new lines executed after the initial seed collection. Note that \restler is run for $48$h in total, but no new lines are discovered. \pythia performs best w.r.t.~ all the baselines.
        % \underline{Seed collection:} Run \restler on each API in order
        % to generate seed test case corpuses. The seed collection time is
        % set to $24$h for all APIs expect for ``issues" in which the respective
        % time was extended to $32$h.
        % Within $24$h \restler reached a plateau in all APIs except for issues,
        % where the respective time had to be extended to $32$h due to a later
        % plateau.
        % \underline{Fuzzing:} Use each corpus to perform three individual
        % $24$h fuzzing sessions per API.
        % %: one for each mutation strategy,
        % %including random
        % %byte-level, random tree-level, and guided tree-level mutations.
        % Moreover, let \restler run for $24$h additional hours ($48$h in total).
        % \underline{Comparison:} Show the number of new lines executed after the
        % initial seed collection. Note that \restler is run for $48$h in
        % total but no new lines are discovered, for no API, in the later $24$h
        % shown in the graphs. Whereas, \pythia fuzzers are run for $24$h,
        % mutating the seed corpuses generated by the first $24$h of RESTler
        % traffic, and discover new lines that have not been covered before.
    }
    \label{fig:all-distillation-off}
    \end{minipage}\\[1em]
\end{figure*}

\section{Experimental Setup}
\label{sec:subj}
% We describe the experimental set up of our evaluation including
% the set up of the target service deployments, the fuzzing set up,
% and the configuration and hyper-parameters of
% the ML models used.

%\heading{Target service deployments.}
\subsection{Study Subjects}
\Cref{tab:target-apis} summarizes the APIs tested by \pythia. In total,
we tested 6 APIs of \gitlab~\cite{gitlab-doc}, 2 APIs of
Mastodon~\cite{mastodon}, and 1 API from Spree~\cite{spree}.
First, we test
\gitlab enterprise edition stable version 11-11 through its REST APIs related
to common version control operations. \gitlab is an open-source web service
for self-hosted Git, its
back-end is written in over 376K lines of ruby code using ruby-on-rails, and
its functionality is exposed through a REST API. It is used by more
than $100,000$ organizations, has millions of users, and has currently a 2/3
market share of the self-hosted Git market~\cite{gitlab-statistics}.
We configure \gitlab to use Nginx HTTP web server and $20$ Unicorn rails
workers limited to up to $4$GB of physical memory. We use postgreSQL for
persistent storage configured with a pool of 20 workers and use the default
\gitlab default configuration for sidekiq queues and redis workers.
According to GitLab's deployment recommendations, such configuration
should scale up to 4,000 concurrent users~\cite{gitlab-requirements}.
Second, we test Mastodon, an open-source, self-hosted social
networking service with more than $4.4$M users~\cite{mastodon-statistics}.
We follow the same configuration with \gitlab regarding Unicorn rails
workers and persistent storage. Third, we test Spree, an open-source
e-commerce platform for Rails 6 with over $1$M downloads~\cite{spree}.

\Cref{tab:target-apis} shows the characteristics of the target service APIs
under tests. All target APIs are related to common operation the users of the
corresponding services may do. In principal, the total number of requests in
each API family along with the average number of available primitive value
combinations for each request indicate the size of the state space that needs
to be tested. Furthermore, the existence of path or body dependencies, or both,
among request types, capture another qualitative property indicative of how
difficult it is to generate sequences of request combinations.

\subsection{Monitoring Framework \& Initial Seed Generation}
We statically analyze the source code of each target service to extract basic
block locations and configure each service, using Ruby's {\tt Class:TracePoint}
hooks, to produce stack traces of lines of codes executed during testing.
During testing, all target services are being monitored by \pythia's coverage
monitor which converts stack traces to bitmaps of basic block activation
corresponding to the test cases executed.
In order to perform test suite minimization (seed distillation), we statically
analyze the source code of each target service and extract  $11,413$ basic
blocks for \gitlab, $2,501$ basic blocks for \mastodon, and $2,616$ basic
blocks for Spree.

\pythia starts fuzzing using an initial corpus
of seeds generated by \restler, an existing, stateful REST API
fuzzer~\cite{restler}. To produce
these initial seeds, we run \restler for a custom amount of time on each
individual API family of each target service using its
default fuzzing mode (\ie Breadth First Search), its default fuzzing dictionary
(\ie two values for each primitive type), and by turning off its Garbage
Collector (GC) to obtain more deterministic results.

% \heading{Seq2seq models.}
% We use an off-the-shelf seq2seq RNN
% with input embedding, implemented in tensorflow~\cite{tensorflow}.
% The model has one layer of  $256$ Gated Recurrent Unit (GRU) cell in the encoder
% as well as in the decoder. Dynamic input unrolling is performed using
% {\tt tf.nn.dynamic} RNN APIs and the encoder is initialized with a zero state.
% We train the model by minimizing the weighted cross-entropy loss for sequences
% of logits using the Adam optimizer~\cite{adam}. We use batches of $32$
% sequences, iterate for $2000$ training steps with a learning rate of $0.001$,
% and an initial embedding layer of size $100$. The vocabulary of the model
% depends on the number of production rules in the fuzzing grammar of each API
% family and ranges in couple of hundred of production rules. Similarly, the
% length of each sequence depends on the specific API and ranges from
% 500 to 780 rules.
% %
% % [VA]: specify the APIs.
% %
% Training such a model configuration
% in a CPU-only machine takes no more than two hours.
% All the experiments discussed in our evaluations were run on Ubuntu 18.04
% Google Cloud VMs~\cite{google-cloud} with 8 logical CPU cores
% and 52GB of physical memory.
% %, without any Graphics Processing Unit (GPU) acceleration.
% Each fuzzing client is used to test a target
% service deployment running on the same machine.

\subsection{Evaluating \pythia}

\subsubsection{Baselines.}
\label{sec:baselines}
We evaluate \pythia
%'s mutation strategy
%w.r.t.
%~three  baselines: (i) \restler, (ii) random byte-level mutation strategy, and
%(iii) random tree-level mutation strategy.
against three baselines.

\noindent
(i) \heading{\restler.} We use \restler both for seed test case generation and
for comparison. On each target API, we run \restler for $2$ days. The first
day, {\em seed collection phase}, is used to generate seed test cases. The
second day, {\em fuzzing phase}, is used for comparison.
%between \pythia and
%\restler.
We compare the incremental coverage achieved by \restler versus
\pythia over the coverage achieved by the initial seed test cases.

\noindent
(ii) \heading{Random byte-level mutations.}
This is the simplest form of mutations.
As suggested by their name, byte-level mutations are random alternations on the
bytes of each seed test case. In order to produce byte-level mutations,
the mutation engine selects a random target position
within the seed sequence and a random byte value (in the range $0-255$),
and updates the target position to the random byte value.
Naturally, this type of mutations are usually neither syntactically nor
semantically valid (defined in~\S\ref{sec:motivation}).

\noindent
(iii) \heading{Random tree-level mutations.}
%The second type of mutations is random, tree-level mutations performed on the leaves of the tree representations of each seed test case.
In order to
produce random tree-level mutations, the mutation engine selects a random
leaf of the respective tree representation and a random rule in $\mathcal{G}$
with a terminal symbol, and flips the target leaf with using the random rule.
The mutations are exclusively performed on the tree leafs, and not in internal
nodes, in order to maintain the syntactic validity of each test case.
However, since the target leafs and the new rules (mutations) are selected
at random for each test case, the target state space for mutations
on realistic tests cases is quite large.
For example, the test case show in~\F\ref{fig:testcase} is represented to a tree
that consisting of 73 leaf nodes and the RG used to produce it has 66 rules with
terminal rules. This, defines a state space of almost 5,000 feasible mutations
only for one seed --- let alone the total size of the state space defined by the
complete corpus of seeds. Next, we evaluate \pythia's learning-based mutation
strategy which considers the intrinsic structure of each test case and
significantly prunes the size of the search space.

\begin{table}[t]
{
    \small
    \begin{center}
        %\begin{tabular}{p{1.2cm} p{2.5cm} p{1.22cm} p{2cm}}
        \begin{tabular}{llrr}
            \toprule
            %{\bf Target Service}            &   {\bf API Family}        &{\bf Total Requests}
            %&{\bf Request Dependencies}\\
            {\bf Target }   &   {\bf API }    & {\bf Total }     & {\bf Request }\\
            {\bf  Service}  &   {\bf  Family} & {\bf  Requests}  & {\bf  Dependencies}\\
            \midrule
            \multirow{1}{*}{\bf GitLab}
                &   Commits                 & 15~(*11)                  & Path, Body\\
                &   Brances                 & 8~~(*2)                   & Path\\
                &   Issues \& Notes         & 25~(*20)                  & Path\\
                &   User Groups             & 53~(*2)                   & Path\\
                &   Projects                & 54~(*5)                   & Path\\
                &   Repos \& Files          & 12~(*22)                  & Path\\
			\midrule
            \multirow{1}{*}{\bf Mastodon}
                &   Accounts \& Lists       & 26~(*3)                    & Path, Body\\
                &   Statuses                & 18~(*19)                   & Path\\
			\midrule
            \multirow{1}{*}{\bf Spree}
                &   Storefront Cart         & 8~(*11)                    & Path\\
            \bottomrule
        \end{tabular}
        \caption{{\bf Target service APIs.}
            Shows number of
            distinct request types in each API family, (*) average number
            of primitive value combinations that are available for each request
            type, and the respective request dependencies.
        }
        \label{tab:target-apis}
    \end{center}
    \vspace{-5pt}
}
\end{table}

\subsubsection{Evaluation}
\label{sec:evaluation}

We answer the following questions:
%obtained with \pythia that answer the following questions:
\begin{enumerate}
    \item[\underline{{\bf RQ1:}}] How do the three baselines
        %(\restler, random
        %byte-level mutations, and random tree-level mutations)
        compare with
        \pythia in terms of code coverage increase over time?
        (Section~\ref{sec:evaluation:mutations})
   % \item[\underline{{\bf RQ2:}}] How does the volume of initial seeds impacts
    %    the code coverage achieved by \pythia?
    %    (Section~\ref{sec:evaluation:seeds-volume})
    \item[\underline{{\bf RQ2:}}] How does initial seed selection impact the code coverage achieved by \pythia?
        (Section~\ref{sec:evaluation:seeds-volume})
    % \item[\underline{{\bf RQ3:}}] What is the code-coverage benefit offered by
    %     test suite minimization (seed distillation)?
    %     (Section~\ref{sec:evaluation:distillation})
    \item[\underline{{\bf RQ3:}}] What is the impact of seed distillation (test suite minimization) on code-coverage?
        (Section~\ref{sec:evaluation:distillation})
    %\item[\underline{{\bf RQ4:}}] How do our coverage results generalize
    %    across all three services and there are any new bugs found?
    %    (Section~\ref{sec:evaluation:generalization-and-bugs-found})
    \item[\underline{{\bf RQ4:}}] Can Pythia detect bugs across all three services?
            (Section~\ref{sec:evaluation:generalization-and-bugs-found})
\end{enumerate}
% We answer RQ1, RQ2, and RQ3 using \gitlab, and answer RQ4 across all three cloud
% services shown in \Cref{tab:target-apis}.

\section{Results}
\label{sec:results}

\begin{figure*}[!htpb]
    \centering
    \subfigure{
        \includegraphics[width=0.3\textwidth]{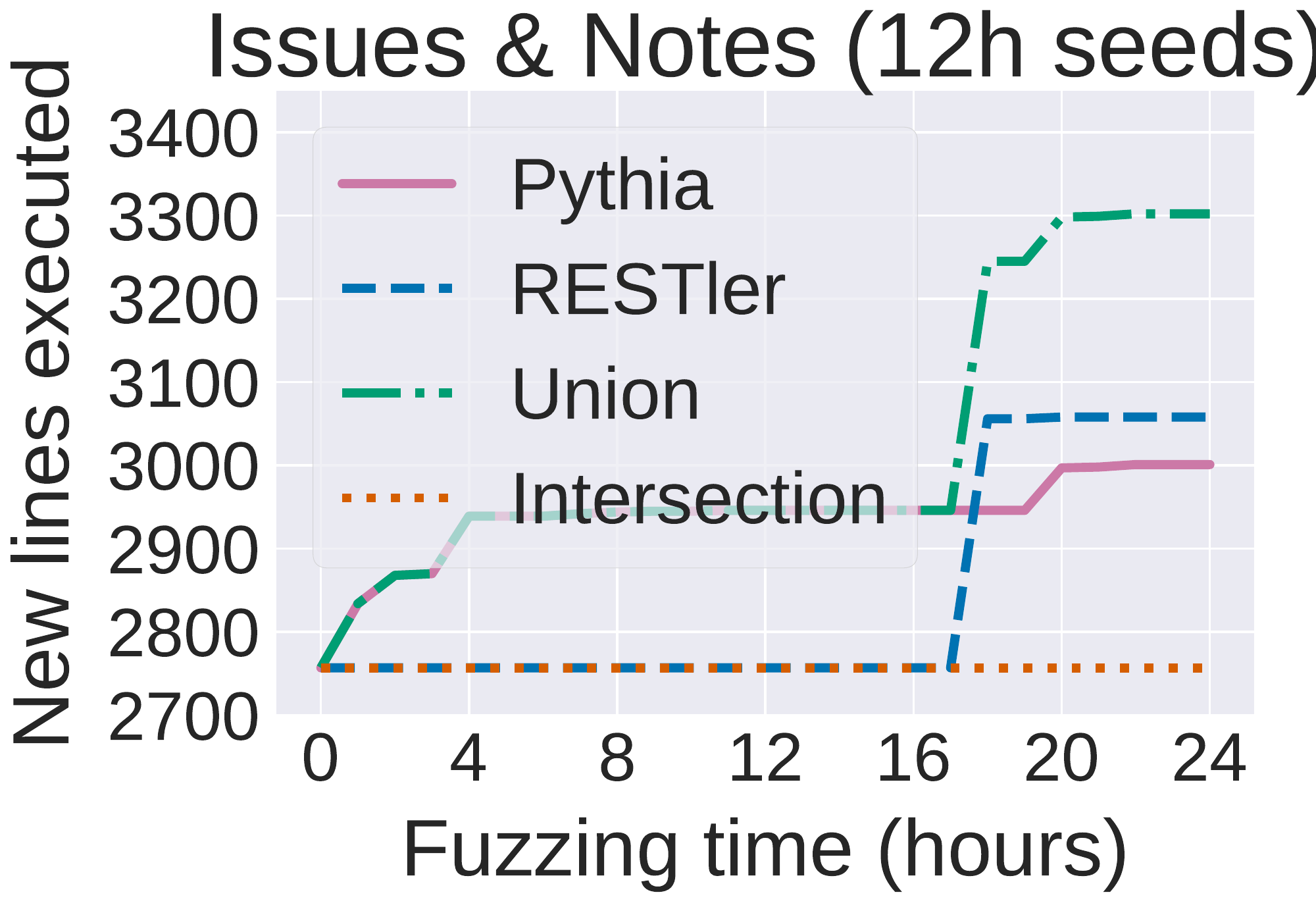}
    }
    \subfigure{
        \includegraphics[width=0.3\textwidth]{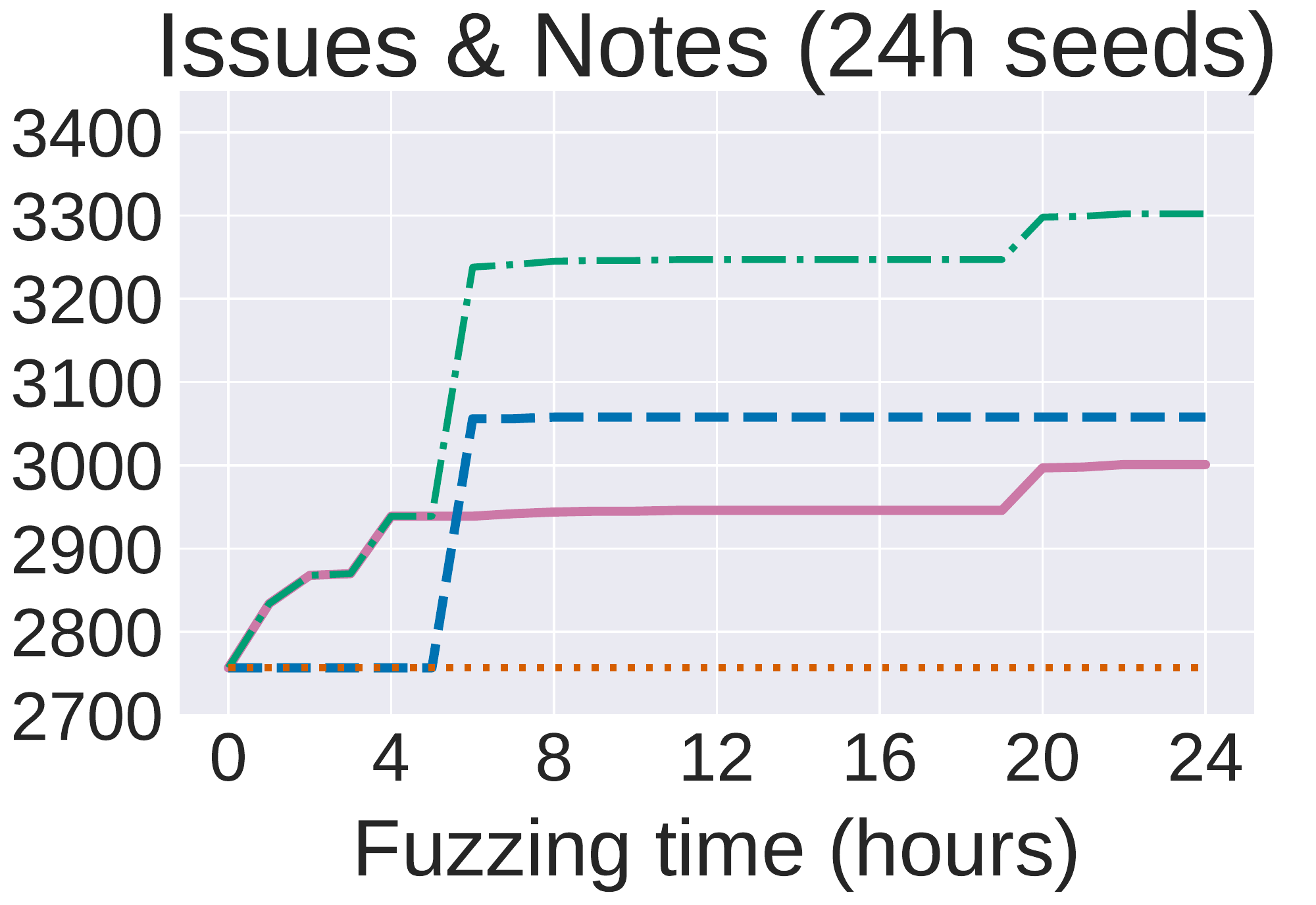}
    }
    \subfigure{
        \includegraphics[width=0.3\textwidth]{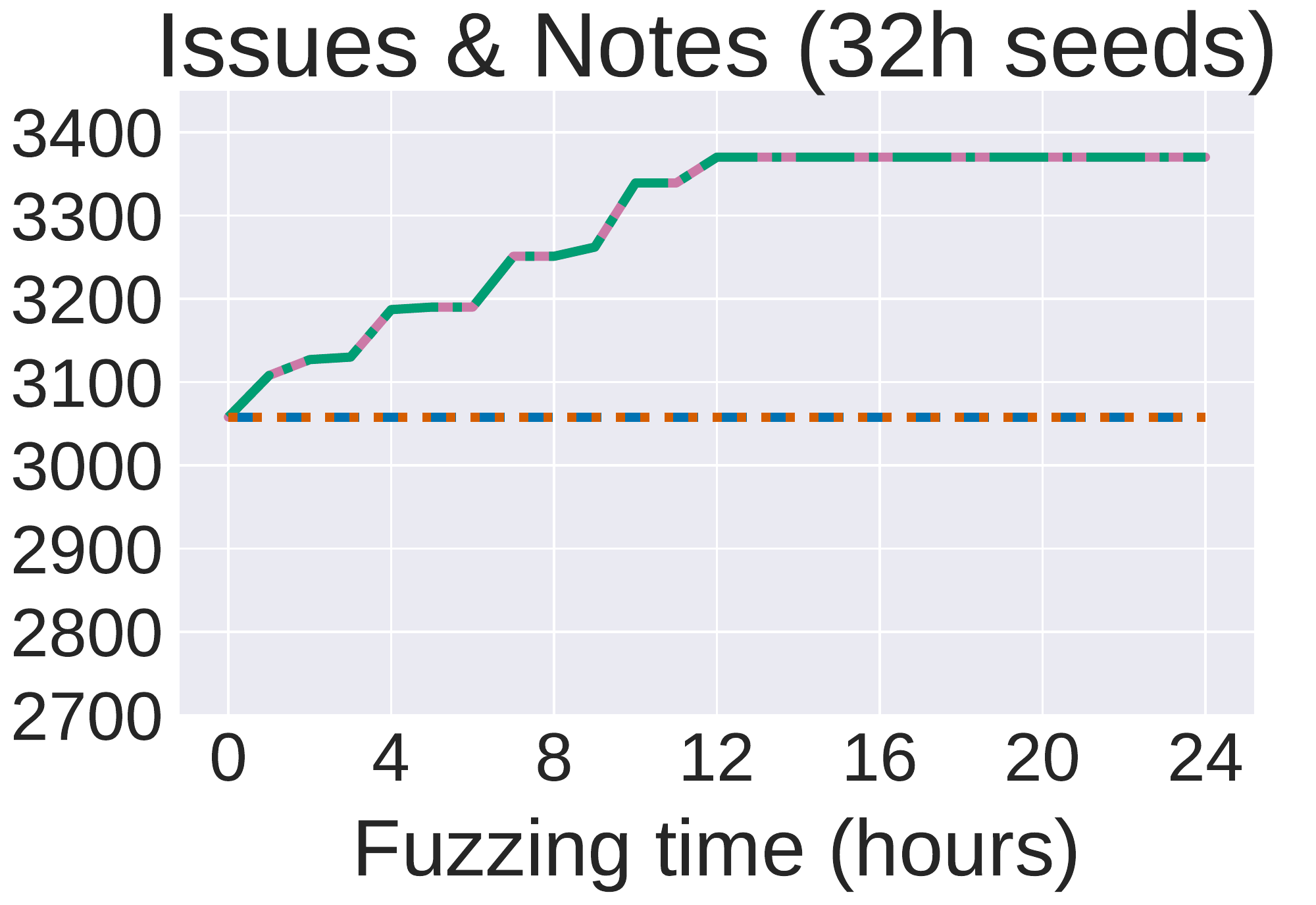}
    }
    \caption{
        {\bf Impact of initial seed collection.}
        \underline{Seed collection:} Run \restler for $12$h on each API.
        \underline{Fuzzing:} Use each corpus to perform three individual
        $24$h guided tree-level \pythia mutation sessions.
        Moreover, let \restler run for $24$h additional hours ($32$h in total).
        \underline{Comparison:} Show the number of new lines executed after the
        initial $12$h of seed collection.
    }
    \label{fig:issues:multiple-seed-volumes}
    \vspace{-3mm}
\end{figure*}

%\subsection{RQ1.~Mutation Strategies}
\subsection{RQ1.~Code Coverage achieved by \pythia}
\label{sec:evaluation:mutations}

In this RQ, we investigate \pythia's impact on the total line
coverage achieved across all the APIs shown in~\Cref{tab:target-apis}.
We compare \pythia against the three baseline fuzzers introduced
in~\Cref{sec:baselines}. In particular, we check whether \pythia can find new lines once \restler reaches a plateau.
We run \restler for $24$h per API setting except  ``Issues \& Notes'', in which the seed collection phase is extended to $32$h due to a late plateau (explained  in \S\ref{sec:evaluation:seeds-volume}). We train \pythia with these seeds and then fuzz with the generated new test inputs for additional $24$h.
The other two baselines also use \restler generated seeds,
mutate these seeds using their own strategies, and
fuzz the target program for $24$h.
\Cref{fig:all-distillation-off} shows the results for \gitlab APIs.
%We make the following observations.

First, we observe that for all the APIs, \pythia exercised unique new lines of code during fuzzing. Since \restler has plateaued after the initial $24$h of seed collection ($32$h for ``Issues \& Notes) no new lines are discovered by \restler during the latter $24$h of fuzzing.
This type of plateau, which is usual in fuzzing, is expected in the case of \restler because it has to explore an exponential state space as the number of requests in a test case increases. For example, after the first $24$h in  ``Commits'' API, \restler has to explore a state space defined by $19,027$ sequences of length five and $11$ feasible renderings each, on average, before moving on to sequences of length six. This state-space explosion is similar across all APIs. Moreover, while stuck searching a large search space, \restler uses repeatedly the same fuzzing
values, generating likely-redundant mutations.

Further, across all APIs, both \pythia and the two random
baselines discover new lines of code that were never executed by \restler. This demonstrates the value of continuously attempting new mutation values instead of repeatedly applying a fixed set of ones in different combinations.
Even the trivial random byte-level mutation finds at least $100$ additional lines, on top of those discovered by
\restler, in all cases. Compared to all baselines, \pythia always increases
line coverage the most, ranging from $180$ additional lines
(in ``Groups \& Member'' APIs) to $410$ extra lines (in ``Commits'').

We also observe that across all APIs the relative ordering of
\pythia and the three baselines remains consistent over time: \pythia performs better than the random tree-level baseline, which, in turn, performs better than the random byte-level baseline. Such ordering is expected. As explained in~\S\ref{sec:evaluation:distillation}, and also motivated in \F\ref{fig:testcase} with a concrete example, raw byte-level mutations tend to violate both semantic and syntactic validity of the seed test cases and consequently underperform compared to tree-level mutations that obey syntactic validity.
Although the latter %random tree-level baseline
produces syntactically valid mutations,
it mutates without any guidance and thus, cannot target its mutation effort to the right direction %using values
that can have larger impacts on the code coverage.
In contrast, \pythia learns the potential mutation location and values from the existing seed corpus and thus increase line coverage faster and higher.

We ran the same experiments across the APIs of Mastodon and Spree
and observed that the relative comparison between \pythia and \restler always yield the same conclusion: overall, \pythia always finds test cases that execute new, additional lines of code not executed by \restler. Specifically, after $24$h of fuzzing,
\pythia finds $48$ new lines in ``Accounts \& List'' and $34$ new lines in ``Statuses'' of Mastodon;
and $214$ new lines in Spree's ``Storefront Chart''.

% The above observations show that \pythia adds value on top of \restler,
% when the later has plateaued. However, it does not shed light into how \pythia
% performs, compared to \restler, before the later plateaus. We discuss this next.

\subsection{RQ2.~Impact of Seed Selection}
\label{sec:evaluation:seeds-volume}

Previously, we saw that well after \restler plateaus, \pythia still discovers
test cases that increase code coverage. However, it is unclear how the two tools compare before \restler plateaus. This leads to question what is the impact of initial seed selection on the line coverage achieved by \pythia.
We select ``Issues \& Notes'' API, which takes a longer time to plateau among all the APIs (\ie after $32$h), and examine three configurations: initial seeds collected after $12$h, $24$h, and $32$h of \restler run. \F\ref{fig:issues:multiple-seed-volumes} shows the results.

In $12$h and $24$h settings, although \pythia started achieving better coverage, \restler took off after a few hours of fuzzing.
However, once \restler plateaus after $32$h, \pythia keeps on finding new code.
 \F\ref{fig:issues:multiple-seed-volumes}
 shows the union and intersection of the lines discovered by \restler and \pythia to understand whether the two tools are converging or orthogonal in terms
 of discovering new lines. We observe across all the plots of
 \F\ref{fig:issues:multiple-seed-volumes} that the intersection remains constant
 while the union increases. This means that the two tools discover diverging
 sets of lines.
 %In fact, in the first two subplots, both \restler and \pythia discovered new lines
 %and the union is the highest line. This means that the two tools are
 %\vlis{I am not sure whether the following line is important.}
 %Note that, the intersection may start from  different points since the seed
 %generation time varies from $12$ to $24$ to $32$.

%\vlis{I think this section needs some explanation, it might not be very clear to a reader who does not know \restler well.}
%These two observations imply that the two tools are not converging.
As explained
earlier, \pythia finds new lines because it performs mutations with new values,
whereas \restler constantly uses a predefined set of values.
In addition, it is now also clear that \pythia cannot cover lines covered
by \restler test cases. This is because, by construction, the mutations
generated by \pythia exercise new rules and lead to syntactically, and largely
semantically, valid test cases but they are not designed
to mutate the request sequence semantics. In other words, no new request
sequence combinations will be attempted by \pythia.
Instead, \pythia focuses on mutations at the primitive types within individual requests. This limitation
becomes evident when, before the last plateau, \restler increases sequence length
(and covers new lines by producing longer test cases), whereas \pythia
has no means of deriving such test cases. The conclusions drawn
by investigating the impact of the initial seed volume in
``Issues \& Notes'' generalize across all APIs tested so far.

\subsection{RQ3.~Impact of Seed Distillation}
\label{sec:evaluation:distillation}

\begin{figure*}[t]
    \begin{minipage}{\textwidth}
    \centering
    \subfigure{
        \includegraphics[width=0.3\textwidth]{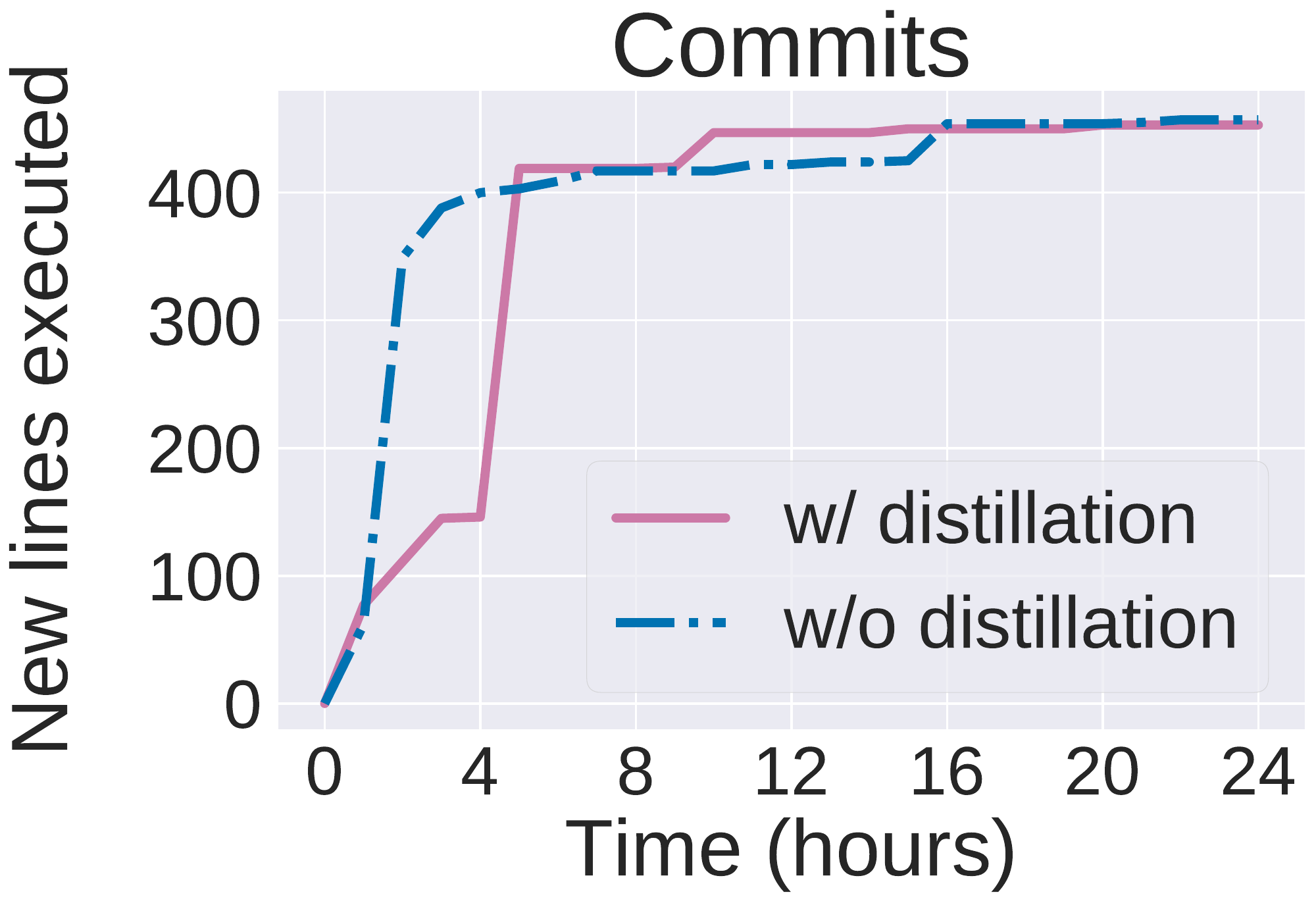}
    }
    \subfigure{
        \includegraphics[width=0.3\textwidth]{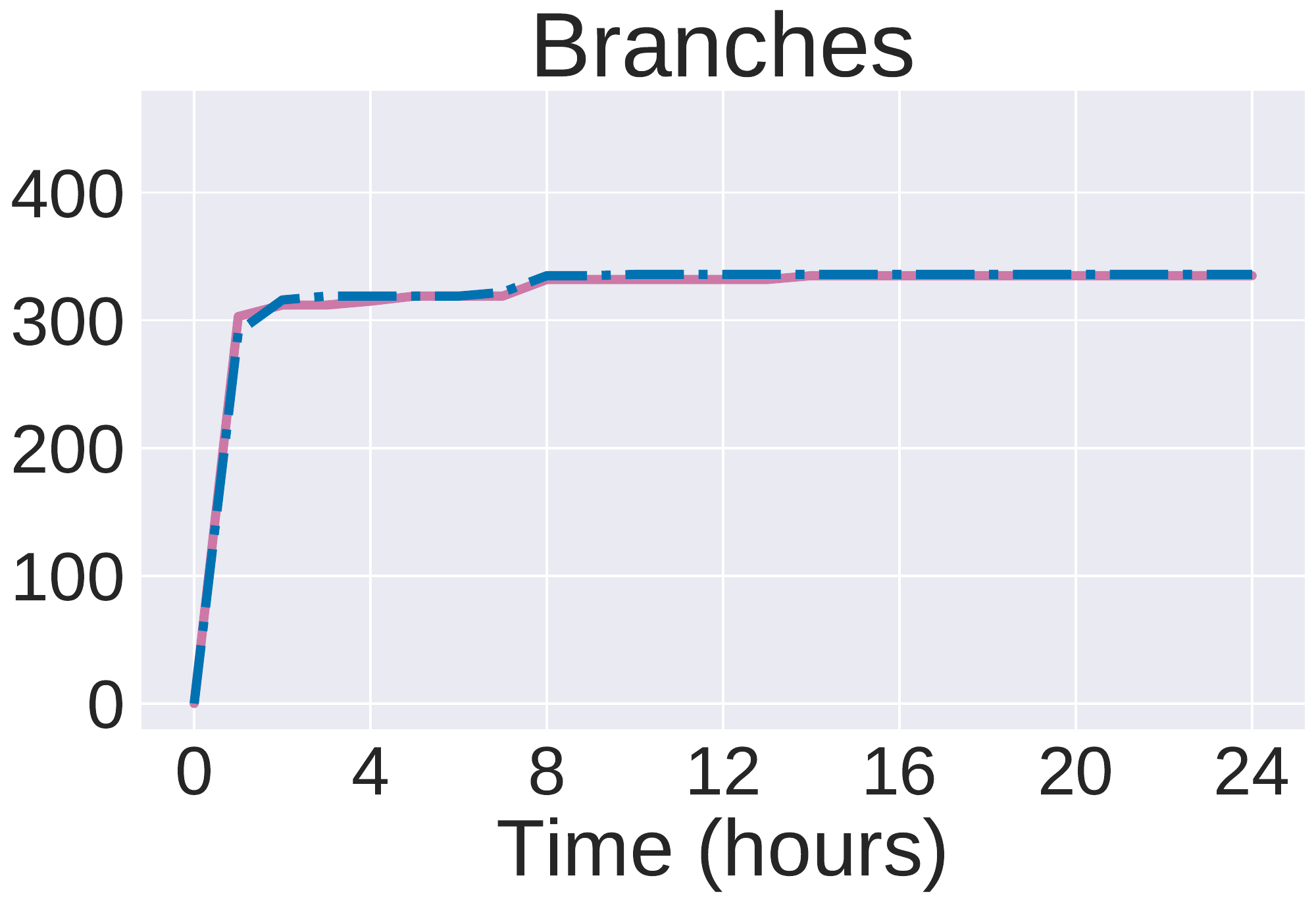}
    }
    \subfigure{
        \includegraphics[width=0.3\textwidth]{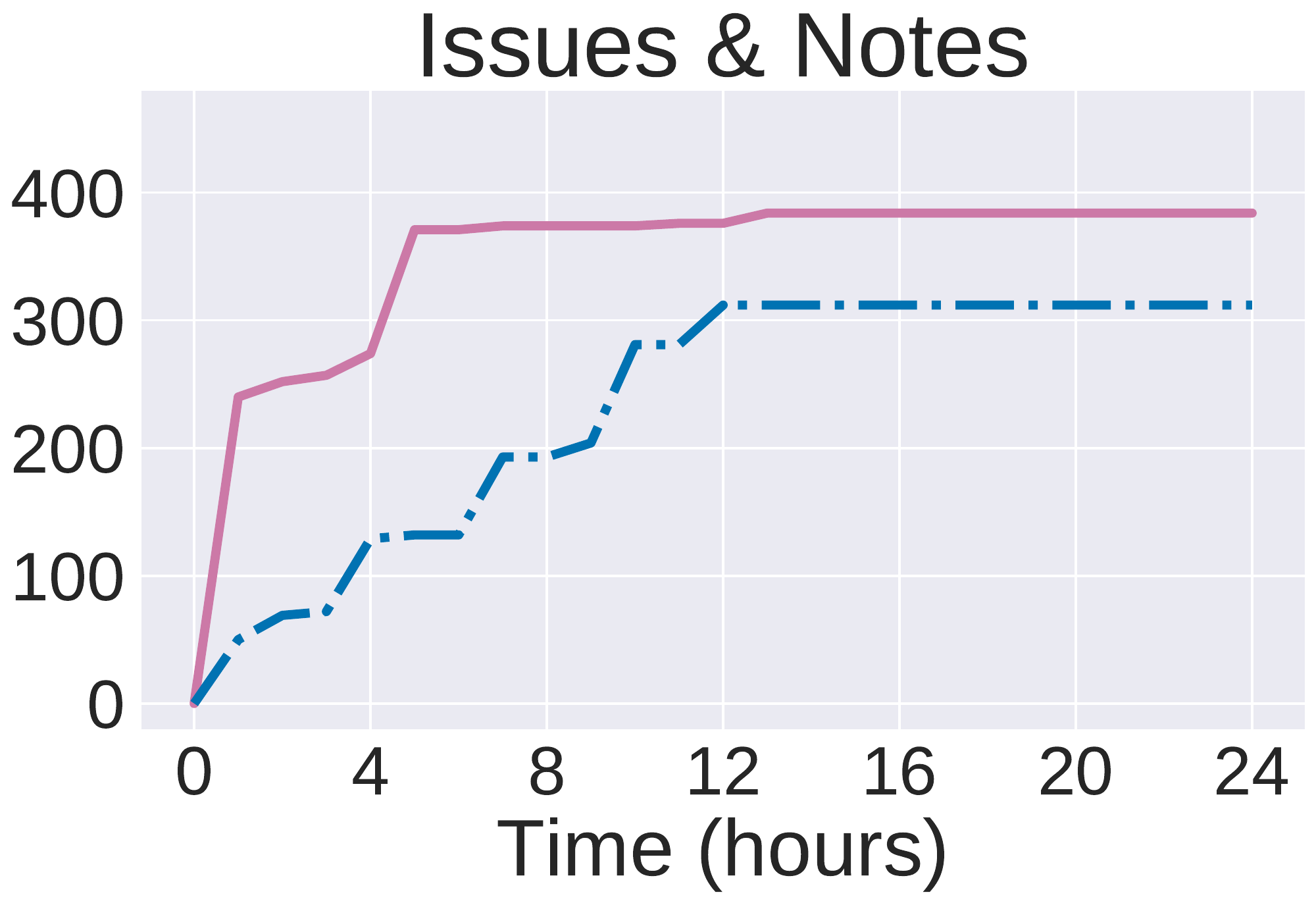}
    }
    %\medskip
    \subfigure{
        \includegraphics[width=0.3\textwidth]{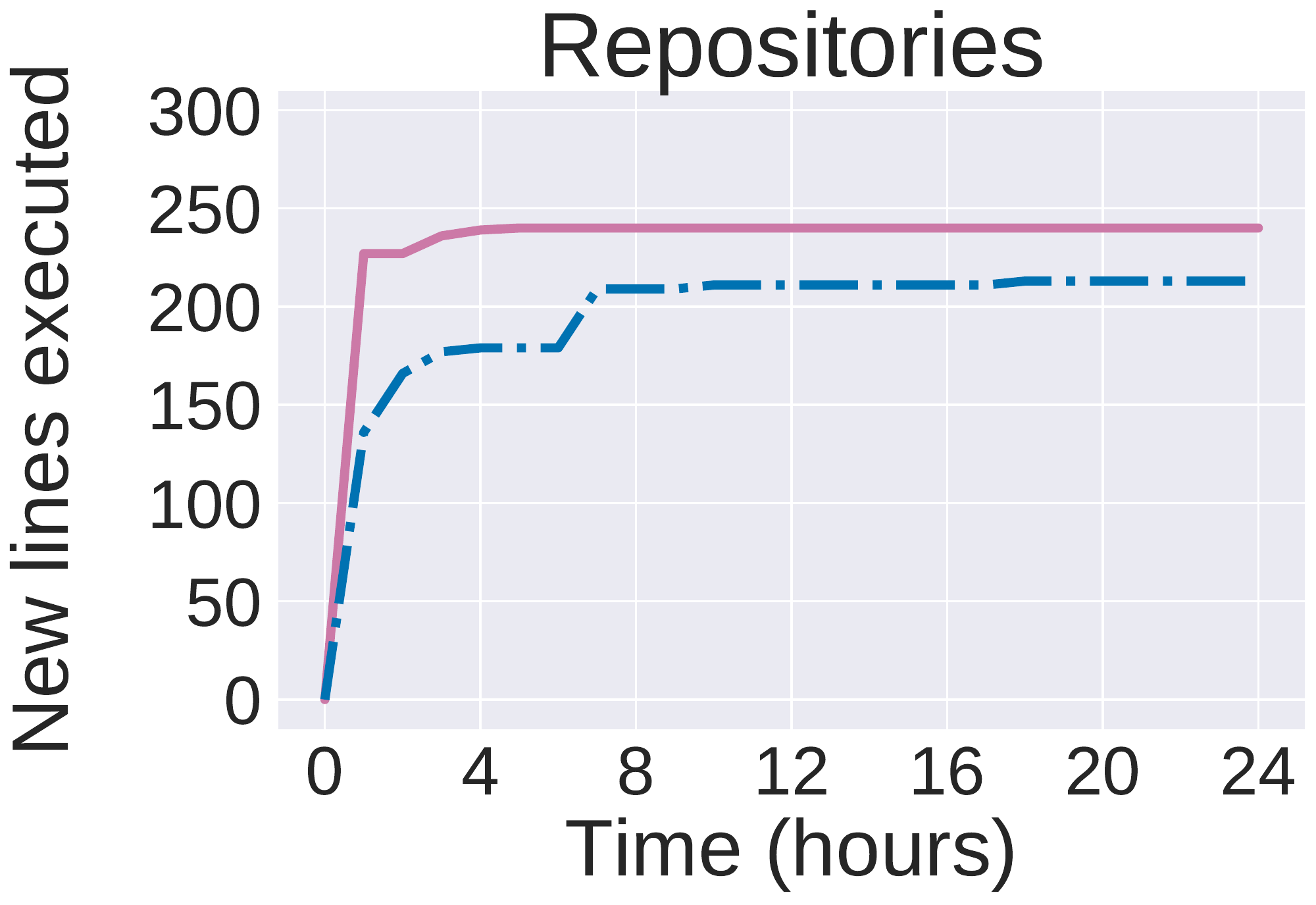}
    }
    \subfigure{
        \includegraphics[width=0.3\textwidth]{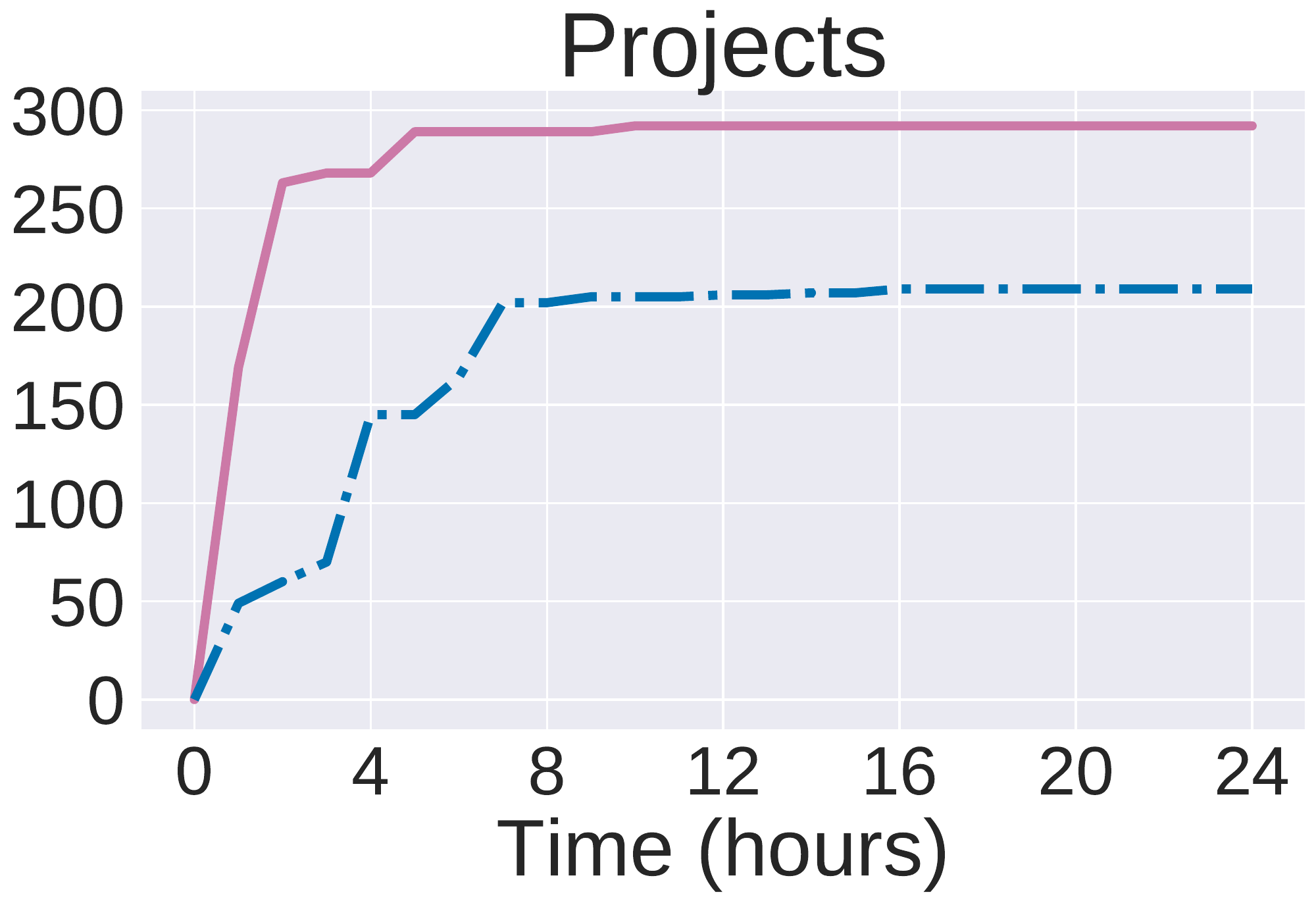}
    }
    \subfigure{
        \includegraphics[width=0.3\textwidth]{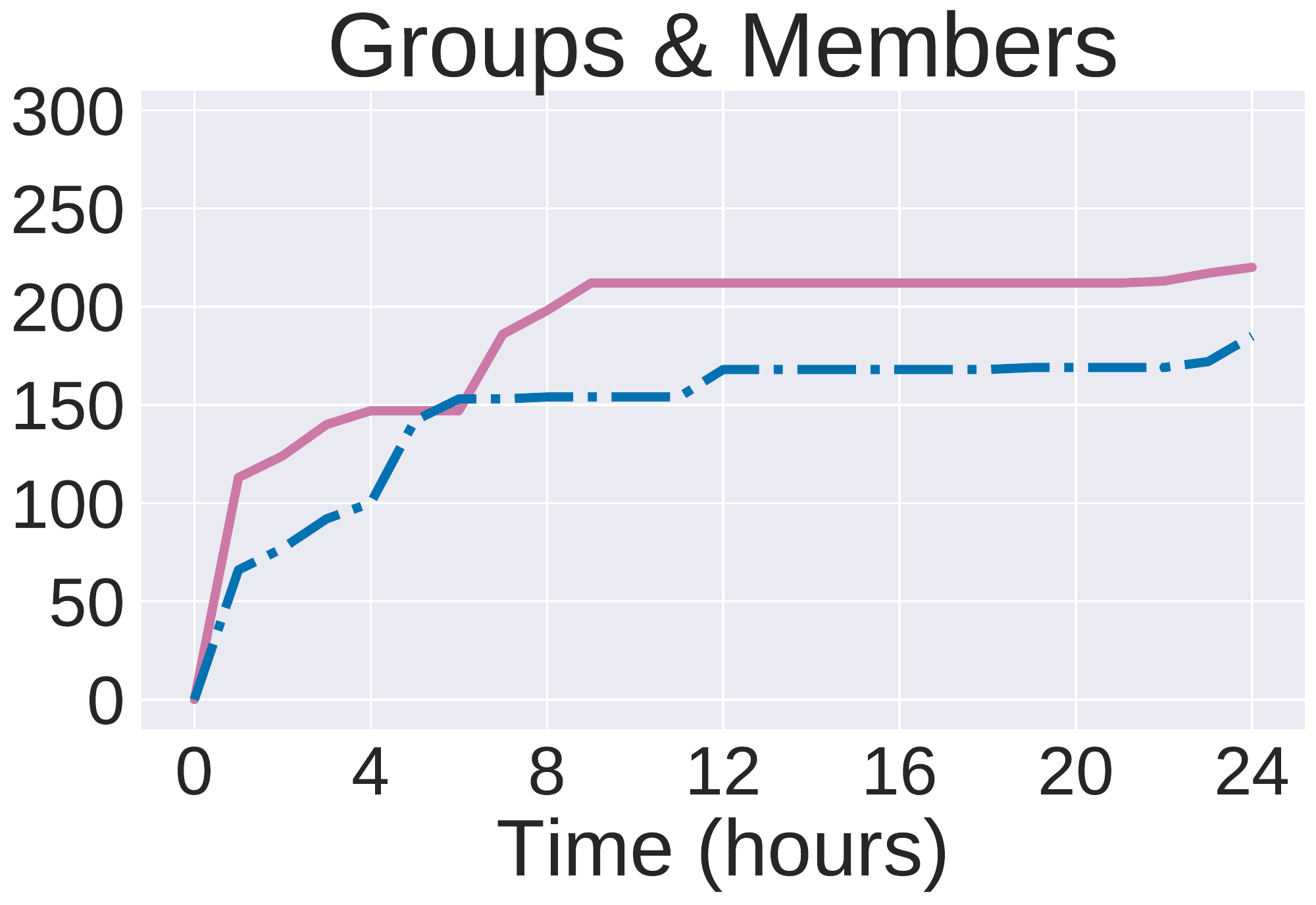}
    }
    \caption{
        {\bf Impact of distillation (test suite minimization).}
        \underline{Seed collection:} Run \restler for  $24$h on each API in order
        to collect seed corpora.
        \underline{Fuzzing:} Use each corpus to perform two
        individual 24h guided tree-level Pythia mutation sessions. One with
        test suite minimization (distillation) and another without.
        \underline{Comparison:} Show the number of new lines executed after
        the initial seed collection.
    }
    \label{fig:all-multitrain}
    \end{minipage}\\[1em]
     \vspace{-5pt}
\end{figure*}
All target services are being monitored by \pythia's coverage
monitor in order to perform test suite minimization, referred to as {\it
seed distillation}. In order to investigate the impact of seed distillation
we perform two independent experiments with and without distillation,
on all \gitlab APIs, using the same initial seeds and fuzzing for $24$h.
\Cref{fig:all-multitrain} shows the total number of additional new
lines executed during the fuzzing phase with and without distillation.

We observe that for most APIs,
%including, ``Issues \& Notes'', ``Repositories'',  ``Projects'', and ``Groups \& Members''
seed distillations help executing more new lines, and faster.
The best incremental benefit is observed in ``Projects'', while the worst
(no benefit at all) is observed for ``Branches''. The ``Branches'' APIs are relatively simple with $8$ total requests and $2$ primitive values.
%combinations on average per request.
In such simple case distillation does not
offer any benefit. Distillation also does not benefit in
``Commits''. Although the setting with distillation outperforms the
one without in the time frame between the fourth and the sixteenth
hour, ultimately, the two settings converge on the same coverage.
%Further hyperparamer tuning on \pythia's seq2seq model or retraining with more
%data (incorporating the new test cases discovered by \pythia) are two potential
%avenues for improvement in such cases.

%!TEX root=../paper.tex
%\subsection{RQ4.~Results Generalization \& Number of Bugs Found}
\subsection{RQ4.~Number of Bugs Found}
\label{sec:evaluation:generalization-and-bugs-found}

% Using \restler seed test cases we, previously, examined the coverage benefit
% obtained by \pythia across all \gitlab APIs. We observed that \pythia consistently
% found test cases that increase the code coverage initially achieved by \restler.
% We ran the same experiments across the APIs of Mastodon and Spree,
% and observed that the relative comparison between \pythia and \restler always
% yield the same conclusion: overall, \pythia always finds test cases that
% execute new, additional lines of code not executed by \restler.

% Specifically, after $24$h of fuzzing, \pythia finds $48$ new lines
% in ``Accounts \& List''
% not found by \restler, $34$ new lines in ``Statuses'', and $214$ new lines
% in ``Storefront Chart''.

Although code coverage is an indicative proxy regarding
the effectiveness of bug finding tools, the ultimate metric is indeed the
total number of bugs found. \pythia found
new bugs across every API and every service tested so far. In total
\pythia found \nbugs new bugs.

While fuzzing with \pythia, there is a high number of ``500 Internal
Server Errors'' received and different instances of the same bugs were reported.
These ``500 Internal Server Errors'' are potential server
state corruptions that may have unknown consequences in the target service
health. Since all the bugs found have to be manually inspected, it is
desirable to report unique instances of each bug and avoid duplication.
To this end, we use the
code coverage information and group bugs using the following rule: {
out of all test cases triggering ``500 Internal Server Error'',
we report those as bugs that are generated by exercising unique code paths.
%from unique code only test cases that both trigger ``500 Internal Server Errors''
%and increase code coverage.
}
%Code coverage increase reflects test
%cases that exercise different code paths and helps avoid duplication.
According
to the aforesaid rule, \Cref{tab:bugs} shows the
bugs found across all services tested. In $24$ hours,
\pythia and \restler generate the same order of magnitude of test cases.
The test cases of both tools have similar execution time in the target services
because the total number of requests per test case remains similar. (\pythia does
not attempt new request sequence combinations.)
However, \pythia's learning-based mutations trigger many more $500$s, which
lead to more unique bugs. \pythia operates on seed test cases generated by
\restler which naturally trigger all bugs found by \restler. We do not count
bugs found by \restler in the results reported for \pythia.
Next, in ~\Cref{sec:case-studies}, we conduct case studies on an indicative
subset of bugs found by \pythia.

\begin{table}[t]
{
    \footnotesize
    \centering
    \def\arraystretch{1.14}
    %\begin{center}
        \begin{tabular}{@{\extracolsep{.09pt}}p{1.8cm}p{0.75cm}p{0.6cm}p{0.6cm}P{0.75cm}p{0.6cm}P{0.6cm}}
		\hline
            \multirow{2}{*}{\parbox{1cm}{\bf Target APIs}}
                & \multicolumn{3}{c}{\restler}
                & \multicolumn{3}{c}{\pythia}
				\\
				\cmidrule(l{8pt}r{8pt}){2-4}
				\cmidrule(l{8pt}r{8pt}){5-7}
                & {\bf Tests} & {\bf 500s} & {\bf Bugs}
                & {\bf Tests} & {\bf 500s} & {\bf Bugs}
			\\
			\hline
		\hline
            Commits                                & $11.6$K 	& $0$ 		& 0& $10.7$K 	& $132$ 	& 3 \\
            Branches                               & $10.3$K 	& $0$ 		& 0& $12.3$K 	& $135$ 	& 4 \\
            Issues                                 & $15.2$K 	& $0$ 		& 0& $11.1$K   	& $246$ 	& 5 \\
            User Groups                            & $9.4$K  	& $0$   	& 0& $15$K 		& $234$ 	& 4 \\
            Projects                               & $11$K 	 	& $0$   	& 0& $18.3$K 	& $185$ 	& 4 \\
            Repos~\&~Files                         & $16.4$K 	& $0$ 		& 0& $14.9$K 	& $79$  	& 2 \\
		\hline
            Accounts~\&~Lists                      & $48.1$  	& $0$   	& 0& $63.5$K 	& $1307$	& 3 \\
            Statuses                               & $58.8$  	& $336$ 	& 1& $56$K 		& $962$ 	& 1 \\
		\hline
            Storefront~~~Cart                      & $15.5$  	& $2018$	& 1& $18.7$K 	& $401$ 	& 3 \\
		\hline
		\hline
            {\bf Total}                            & - 			& - 		& 2	&  -  		&- & {\bf \nbugs}\\
		\hline
        % END: This is automatically generated
        \end{tabular}
        \caption{Number of test cases generated, ``500 Internal Server Errors'' triggered,
			and unique bugs found by \restler and \pythia after $24$h of fuzzing.
        }
       \label{tab:bugs}
       \vspace{-10pt}
       %\end{center}
}
\end{table}

\section{New Bugs Found}
\label{sec:case-studies}
During our experiments with \pythia on local \gitlab,
\mastodon, and Spree deployments we found $\nbugs$ new bugs. All bugs were
easily reproducible and we are in the process of reporting them to the respective
service owners. We describe a subset of those bugs to give a flavor of what
they look like and what test cases uncovered them.

\heading{Example 1: Bug in Storefront Cart.}
One of the bugs found by \pythia in Spree is triggered when a user tries to
add a product in the storefront cart using a malformed request path
\linebreak
{\tt ``/storefront/|add\_item?include=line\_items''}.
Due to erroneous input sanitization,
the character {\tt ``|''} is not stripped from the intermediate path parts.
 Instead, it reaches the function {\tt split} of library
{\tt uri.rfc3986\_parser.rb}, which treats it as a delimiter of the path string.
This leads to an unhandled {\em InvalidURIError} exception in the caller
library {\tt actionpack}, and causes a ``500 Internal Server Error''
preventing the application from handling the request and
returning the proper error, \ie ``400 Bad Request''. This bug can be reproduced
with a test case with two requests: (1) creating a user
token and (2) adding a product in the chart using a malformed request path.
Bugs related to improper input sanitization and unhandled values passed
across multiple layers of software libraries are usually found when using
fuzzing. \pythia found bugs due to malformed request
paths in all the services tested.

\heading{Example 2: Bug in Issues \& Notes.}
Another bug found by \pythia in \gitlab's Issues \& Notes APIs is triggered when
a user attempts to open an issue on an existing project, using a malformed
request body. The body of this request includes
multiple primitive types and multiple key-value pairs, including
{\tt due\_date, description, confidentiality,
title, asignee\_id, state\_event}, and others.
A user can create an issue using a malformed value for the field title, such as
{\tt \{"title":"DELE\textbackslash xa2"\}}  which leads to a
``500 Internal Server Error''. The malformed title value is not sanitized
before reaching the fuction {\tt create} of {\tt <class:Issues>} that
creates new issues. This leads to an unhandled
{\tt ArgumentError} exception due to an invalid UTF-8 byte sequence.
This bug can reproduced by (1) creating a project and (2) trying to post an
issue with a malformed title in the project created in (1).

Interestingly,
adding malformed values in other fields of the request body does not
necessarily lead to errors. For instance, the fields {\tt
confidentiality} and {\tt state\_event} belong to different primitive types
(boolean and integer) which are properly parsed and sanitized.
Furthermore, mutations that break the json structure of the request body or
that do not use an existing project ids also do not lead to such errors.
Brute-forcing all possible ways to break similar REST API request sequences
is infeasible. Instead, \pythia
learns common usage patterns of the target service APIs and then applies
learning-based mutations breaking these common usage patterns, while still maintaining
syntactic validity.
\pythia found such input sanitization bugs, due to malformed request bodies,
in all services tested.
Similar bugs are shown
in~\Cref{fig:testcase,fig:mutationInternal,fig:mutationExternal}.

Other examples of unhandled errors found by \pythia are due to
malformed headers and request types. All the bugs found in this work
are currently being reported to the service owners.

%Although we are not the owners of the tested services, such error appear
%less likely to permanently harm the state of the target services, since they do
%not seem to transfer across libraries that touch persistent storage or
%the key-value stores. Nevertheless, we are in the process of reporting all bugs
%found for the respective service owners to judge.

\section{Related work}
\label{sec:related-work}
Our work aims at testing cloud services with REST APIs
and relates with works across three broad domains: (i) blackbox
grammar-based fuzzing, (ii) coverage-guided and fully-whitebox fuzzing
approaches, and (iii) learning-based fuzzing approaches.

In blackbox grammar-based approaches, the user provides an input grammar
specifying the input format, what input parts are to be fuzzed and
how~\cite{boofuzz,burp,Peach,SPIKE}. A grammar-based fuzzer then generates
new inputs, satisfying the constraints encoded by the grammar. These
new inputs reach deep application states and find bugs beyond syntactic lexers
and semantic checkers. Grammar-based fuzzing has recently been
automated in the domain of REST APIs by \restler~\cite{restler}.
\restler performs a lightweight static analysis of the API specification
in order to infer dependencies among request types, and then automatically
generates an input grammar that encodes sequences of requests
in order to exercise the service more deeply, in a stateful manner.
\restler  inherits two of the typical limitations of grammar-based
fuzzing, namely fuzzing rules with predefined sets of values and lack of
coverage feedback. \pythia addresses these limitations and augments
blackbox grammar-based fuzzing with coverage-guided feedback and a
learning-based mutation strategy in the domain of stateful REST API fuzzing.

Fuzzing approaches based on code-coverage feedback~\cite{AFL}
are particularly
effective in domains with simple input formats
but struggle in domains with complex input formats. Fully whitebox approaches
can be used to improve test-generation precision by leveraging sophisticated
program analysis techniques like symbolic execution, constraint
generation and solving~\cite{DART,EXE,klee,SAGE,anand2007jpf,avgerinos2014enhancing,chipounov2012s2e}
but still fall short of grammar-based fuzzing when required to generate
syntactically and semantically valid inputs.
As an alternative, coverage-guided feedback and domain-specific heuristics
asserting generation of semantically valid inputs have been
combined~\cite{padhye2019semantic,pham2019smart}.
More heavy-weight whitebox fuzzing techniques~\cite{MX07,GKL08,rawat2017vuzzer}
have also been combined with grammar-based fuzzing.
All these approaches are not learning-based. 
In contrast, \pythia uses a learning-based approach
and utilizes initial seeds to learn common usage patterns which are then mutated
while maintaining syntactic validity.

Learning-based approaches have recently been used in fuzzing for statistical modeling of test inputs
~\cite{godefroid2017learn,wang2017skyfire}
and for generating regular or context-free input
grammars~\cite{bastani2017synthesizing,autogram,wang2017skyfire}.
These approaches do not utilize any coverage feedback during fuzzing.
Other learning-based approaches aim at modeling the branching behavior
of the target program~\cite{neuzz,rajpal2017not,bottinger2018deep} using a Neural Network (NN)
model. The trained NNs can then be combined with a coverage-guided fuzzer~\cite{AFL}
and used as a classifier to help avoid executing test inputs that are unlikely
to increase code coverage~\cite{rajpal2017not}. Alternatively, the gradients of
the trained NNs can be used to infer which input bytes should be
mutated in order to cover specific
branches~\cite{neuzz}. However, the trained NNs approximate only a small
subset of all possible program behaviours, and these approaches have been applied
only to domains with relatively simple input structures.

%PG-comment: the text above is a start, but it is not a comprehensive comparison with
% closely-related work. Specifically, the paper should explain what is both common
% and different from, at least, learn-and-fuzz and neuzz (especially given
% the co-authors of this paper). But the other cited related work (some of which I am
% not familiar with) should also be compared in a bit more detail. If we get expert
% reviewers, they might likely cringe at how superficial the last paragraph is.
% Can someone please help here?
\section{Threats to validity}
\label{sect:threats}
%Some threats that affect the validity of our study are:\\
%\noindent{\em Hyperparameters:}
%The success of \pythia depends on the choices of different hyperparameters for
%training the seq2seq autoencoder. We empirically determine the optimal
%parameters to ensure maximum edge coverage.
%
%\noindent{\em Static Analysis}:
%We statically analyze the source code of the target services and extract
%basic block locations which we track during fuzzing. This step is typical on
%coverage-guided fuzzing approaches on native C binaries~\cite{AFL}. However,
%since we analyze interpreted code, we miss a subset of basic
%block that are defined within complex if-else list comprehension structures.
%Yet, we track a significant number of basic block in all targets
%(\eg $11,413$ in \gitlab).
%
%\noindent{\em Generalizibility}:
%We only studied three target programs with nine APIs. However, we targeted
%complex production-scale services, with hundrends of API requests
%(see~\Cref{tab:target-apis}) and we believe our
%results will generalize across cloud services with REST APIs.
Some threats that affect the validity of our study are the following.
First, the success of \pythia depends on the choices of different hyperparameters
for training the seq2seq autoencoder. We empirically determine the optimal
parameters to ensure maximum edge coverage. Furthermore,
the static analysis of the source code of target services to extract
basic block is imprecise. Since we analyze interpreted code, we miss a subset
of basic block that are defined within complex if-else list comprehension structures.
Yet, we track a significant number of basic block in all targets
(\eg $11,413$ in \gitlab). Finally, we only studied three target programs with
nine APIs. Yet, we targeted complex production-scale services, with hundrends
of API requests (see~\Cref{tab:target-apis}) and we believe our
results will generalize across cloud services with REST APIs.

\section{Conclusion}
\label{sec:conclusion}
\pythia is the first fuzzer that augments grammar-based fuzzing with
coverage-guided feedback and a learning-based mutation strategy for stateful
REST API fuzzing. Pythia uses a statistical model to learn common
usage patterns of a REST API from seed inputs, which are all structurally
valid. It then generates learning-based mutations by injecting
a small amount of noise deviating from common usage patterns while still
maintaining syntactic validity.
\pythia's learning-based mutation strategy helps generate grammatically
valid test cases and coverage-guided feedback helps prioritize the
test cases that are more likely to find bugs. We presented detailed experimental
evidence\textemdash
collected across three productions-scale, open-source cloud services\textemdash
showing that \pythia outperforms prior approaches both in
code coverage achieved and, most crucially, in new bugs found.
\pythia found new bugs in all APIs and all services tested so far.
In total, \pythia found \nbugs bugs which we are in the process of reporting to
to the respective service owners.

\bibliographystyle{abbrv}
\bibliography{paper}
\end{document}